\documentclass[5p,11pt]{elsarticle}

\usepackage{lineno}
\usepackage{lipsum}
\usepackage{caption}
\usepackage{subcaption}
\usepackage[hidelinks]{hyperref}
\usepackage{booktabs} 

\title{New Harmonic Coils for Enhanced Beam Extraction at the U-120M Cyclotron}

\author[1,2]{Tomáš Matlocha\texorpdfstring{\corref{cor1}}{}}%
\ead{matlocha@ujf.cas.cz}
\author[1]{Jan Štursa}%
\author[1]{Radek Běhal}%
\author[1]{Václav Zach}%
\author[1]{Milan Čihák}%

\cortext[cor1]{Corresponding author}

\affiliation[1]{organization={Nuclear Physics Institute of the CAS}, addressline={Husinec - Řež 130},postcode={250 68},city={Řež},country={Czech Republic}}

\affiliation[2]{organization={Faculty of Nuclear Sciences and Physical Engineering of CTU}, addressline={Břehová 7},city={Prague},postcode={115 19},country={Czech Republic}}

\date{January 2024}

\begin{document}


\begin{abstract}
Since its commissioning in 1977, the U-120M cyclotron has been equipped with one pair of harmonic coils  dedicated to establishing optimal beam quality at low accelerating radii. In 2022, the cyclotron was upgraded and a second pair of harmonic coils was added. The new harmonic coils are positioned at a radius of approximately 50\,cm corresponding to the outermost beam orbit, where they were installed to ensure  efficient extraction of positively accelerated ions from the cyclotron in the newly developed extraction system. This article describes in detail the design of the new harmonic coils, their installation in the accelerator vacuum chamber, and their use in the extraction process. Furthermore, the design of a new extraction system using the precession and linear integer resonance extraction methods is presented and a comparison of the basic properties of these two methods is given.  Finally, an evaluation of their advantages and limitations concerning their possible use in the cyclotron and an estimation of the efficiency of the new extraction system is made.
\end{abstract}

\maketitle


\section{Introduction}

The U-120M cyclotron, operated at the Nuclear Physics Institute of the Czech Academy of Sciences, is a warm magnet, multi-particle, isochronous cyclotron with a wide energy range. It has magnet pole diameter 120\,cm, $K$-value of up to 40\,MeV, and accelerates particles with a mass-to-charge ratio $A/Z$ of 2 or less. The machine is employed to accelerate helium-3 up to 55\,MeV, alpha particles and protons up to 40\,MeV, and deuterons up to 20\,MeV, achieving intensities of up to several tens of microamperes. The operation of the cyclotron allows for the acceleration of positive or negative ions produced by an internal PIG ion-source. More comprehensive information about the cyclotron facility can be found in Ref. \cite{filip_irad_setup}.


In the last two decades, the range of applications of accelerated beams has expanded significantly, especially the production of radionuclides for the preparation of radiopharmaceuticals in nuclear medicine, and the cyclotron has been in almost continuous operation with approximately 3000 operating hours per year. Furthermore,  the cyclotron has been extensively used in basic research experiments and applications such as measurement and validation of nuclear data for astrophysics \cite{dagata}, production of homologues of superheavy elements \cite{BARTL} (solely for rarely accessible beams of helium-3), measurement of excitation functions of various nuclear reactions \cite{cervenak}, ion-beam irradiation of biological samples \cite{vysin}, testing of new type of dosimeters and cosmic-ray detectors \cite{granja}, assessment of the radiation hardness of electronic components  \cite{filip_irad_setup}, production of fluorescent nanodiamonds \cite{STURSA}, production of calibration sources \cite{aker}, assessment of the damage to proton-irradiated samples (DPA), and fusion relevant neutronics experiments  \cite{STEFANIK2}.
In connection with the developed target stations the cyclotron is a unique and powerful source of high-intense fast neutron beams \cite{STEFANIK}.
The output proton flux spans a wide range and can be smoothly regulated within 15 orders of magnitude \cite{ja2} and, moreover, it can be modulated into ultra-short pulses \cite{buncher}.

\begin{figure}[ht]
\centering
\includegraphics[width=0.47\textwidth]{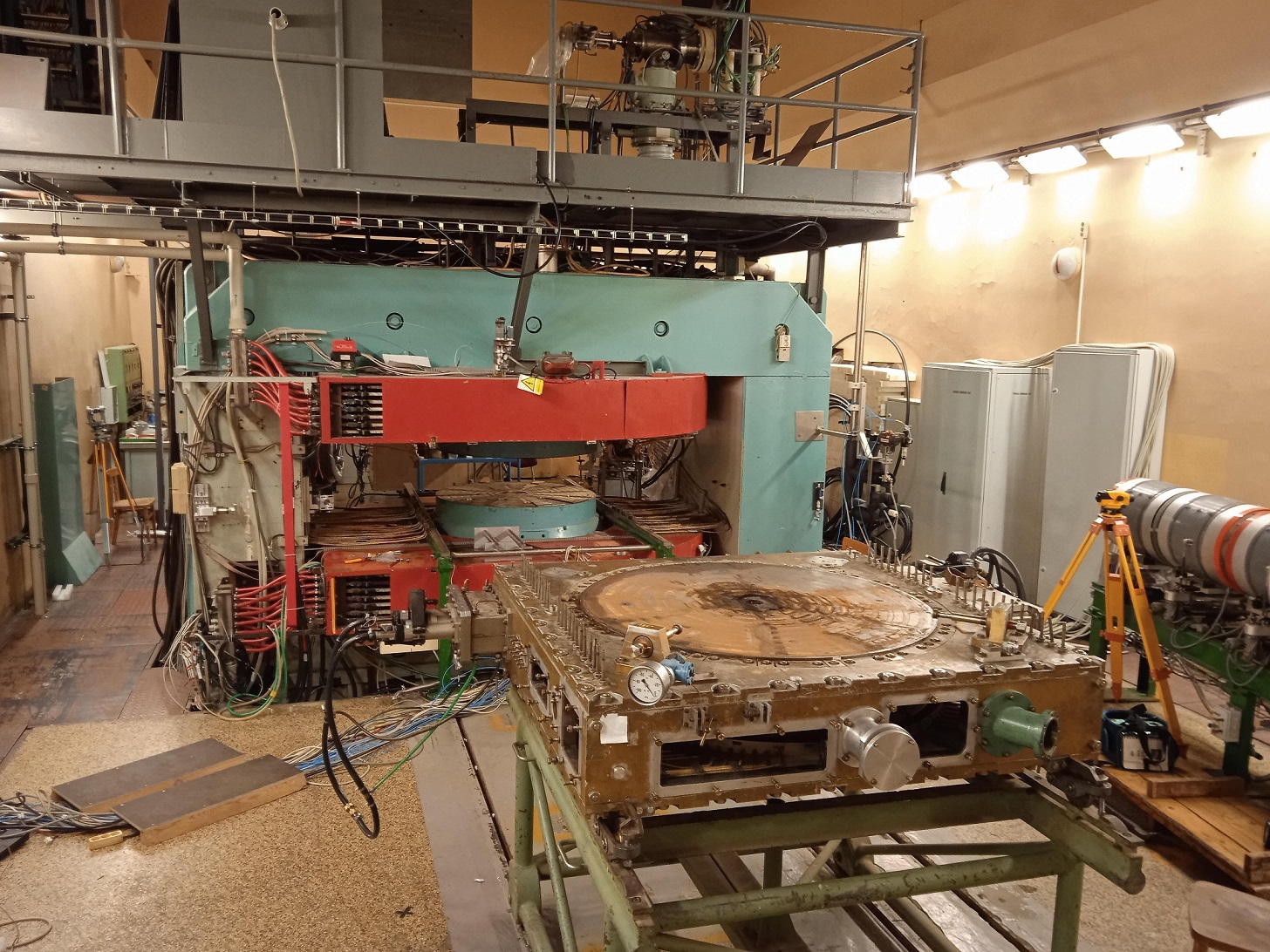}
\caption{Vacuum chamber pulled out from the main cyclotron magnet during the shutdown in 2022. }
\label{fig:magnet_vyjety}
\end{figure}

Since 2015, the operation suffered from permanent short circuit which began to appear on the trim-coil No. 6. In 2022, after the transfer of radionuclide production to the new TR-24 cyclotron \cite{tr24}, it was possible to proceed with the dismantling of the vacuum chamber and start with the complete overhaul of the cyclotron. This forced operational interruption provided an opportunity for a significant modernization of the accelerator, which took a place during a nine-month shutdown. This was the first time since 1988 that the accelerator chamber was moved outside the main magnet region, as shown in Figure~\ref{fig:magnet_vyjety}, and allowed us to address and upgrade the cyclotron's positive-ion extraction system. The extraction efficiency in the positive-ion mode, which ranges from approximately 10\% to 15\% depending on the type and energy of the particles being accelerated, is currently the main drawback of the U-120M cyclotron. Before 2018, when the cyclotron was primarily used for production of radiopharmaceuticals and was mostly operated in the negative-ion mode,  the low extraction efficiency in the positive mode was deemed acceptable. However, given the significantly increasing interest of experimenters in using higher ion currents extracted in the positive mode, the current goal is to improve the corresponding extraction efficiency.

During the 2022 shutdown period, in addition to repairing the damaged trim-coil, new harmonic coils were installed within the volume of the accelerating vacuum chamber. These harmonic coils, positioned on the extraction radius of the accelerator, constitute a fundamental element of the new extraction system.

\begin{figure}[ht]
\centering
\includegraphics[width=0.48\textwidth]{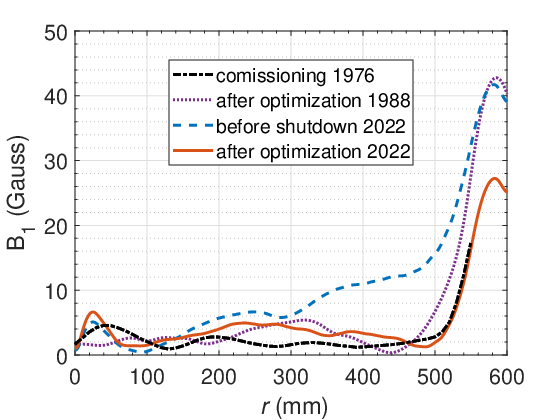}
\caption{Comparison of the individual measurements of the magnetic-field first-harmonic component.  }
\label{fig:first_harm_compar}
\end{figure}

Before repairing the trim-coils, a control measurement of the magnetic field was conducted, revealing a significant first-harmonic perturbation component in the magnetic field. This component was later minimized using a procedure implemented at the AIC-144 cyclotron in Cracow \cite{aic_shift}, which is of a similar type. Employing a very precise horizontal displacement (0.1\,mm) of the acceleration chamber between the magnet poles and monitoring the response in the first-harmonic component,  the first-harmonic was successfully reduced from an initial value of 15\,Gauss that was measured at the beginning of the shutdown at radius 50\,cm to approximately 2\,Gauss obtained by the end of the shutdown in 2022. This final value is comparable to the magnitude of this disturbance during the commissioning of the cyclotron at JINR in 1976. Figure \ref{fig:first_harm_compar} provides a comparison of the first-harmonic component along the radius before and after the shutdown, together with the values optimized in the previous periods.

\begin{figure*}
\centering
\begin{subfigure}[b]{0.49\textwidth}
     \centering
\includegraphics[width=0.88\textwidth]{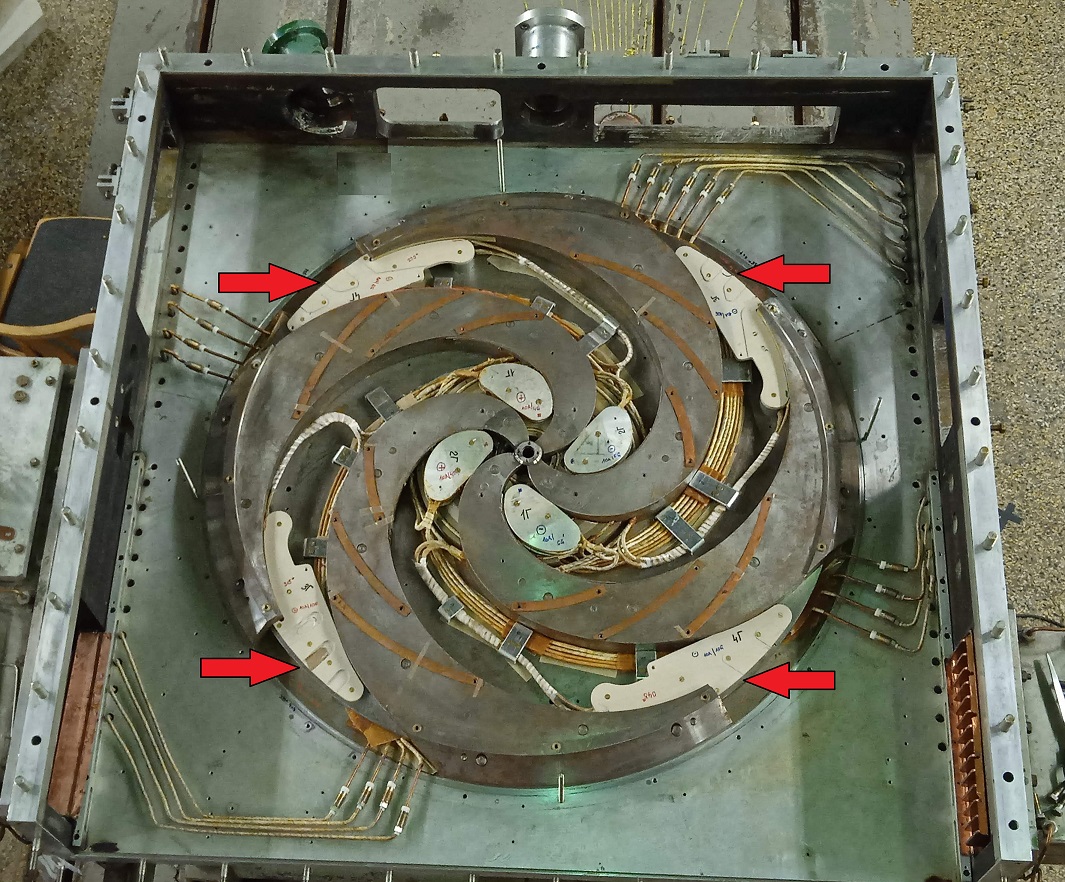}
\caption{Top view of the U-120M bottom sectors. Red arrows indicate the position of the new harmonic coils (OHC).  }
\label{fig:new_HC_in_chamber}
    \end{subfigure} 
\begin{subfigure}[b]{0.49\textwidth}
     \centering
\includegraphics[width=\textwidth]{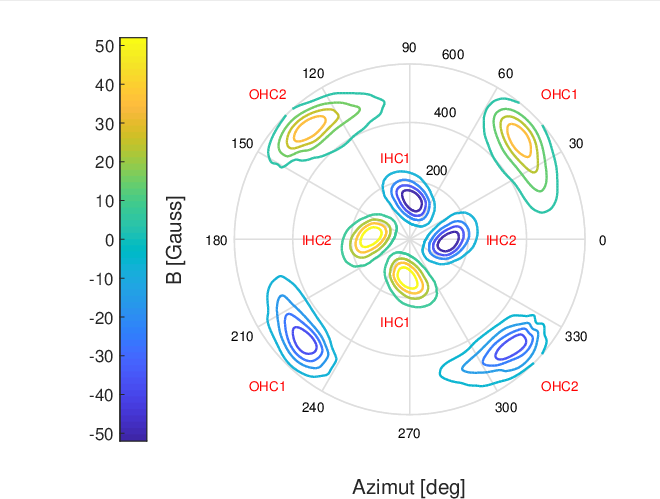}
\caption{Measured magnetic field maps of inner (IHC) and outer (OHC) harmonic coils for 200A current. }
\label{fig:hc_maps}
\end{subfigure}
    \caption{Harmonic coils installed between the cyclotron sectors and their magnetic field maps.}
    \label{fig:HC_coils_full}
\end{figure*}

Figure \ref{fig:HC_coils_full} then illustrates the complete set of harmonic coils installed in the magnetic structure and the magnetic field maps of the inner (IHC) and the newly installed
outer (OHC) harmonic coils. The position of the new harmonic coils (OHC) is indicated by the red arrows. These new harmonics are the basis of a new extraction system whose parameters will be further optimized in the future.

This paper outlines the progress made in developing a new positive-ion extraction system for the cyclotron U-120M. Section \ref{sec:hccoils} presents the design and installation of the new harmonic coils. Subsequently, Section \ref{sec:extraction} provides a description of two selected methods for orbit separation at the extraction radius, along with details on the methods employed in numerical simulations. Section \ref{sec:results} then presents the outcomes of the extraction system simulations, along with optimizations and an evaluation of the extraction efficiency. Section \ref{sec:discussion} is devoted to discussion the obtained results.  Properties of both extraction concepts are compared
and validity of the simulation settings is assessed.


\section{Design and installation of the new harmonic coils}
\label{sec:hccoils}

\begin{figure}[ht]
\centering
\includegraphics[width=0.48\textwidth]{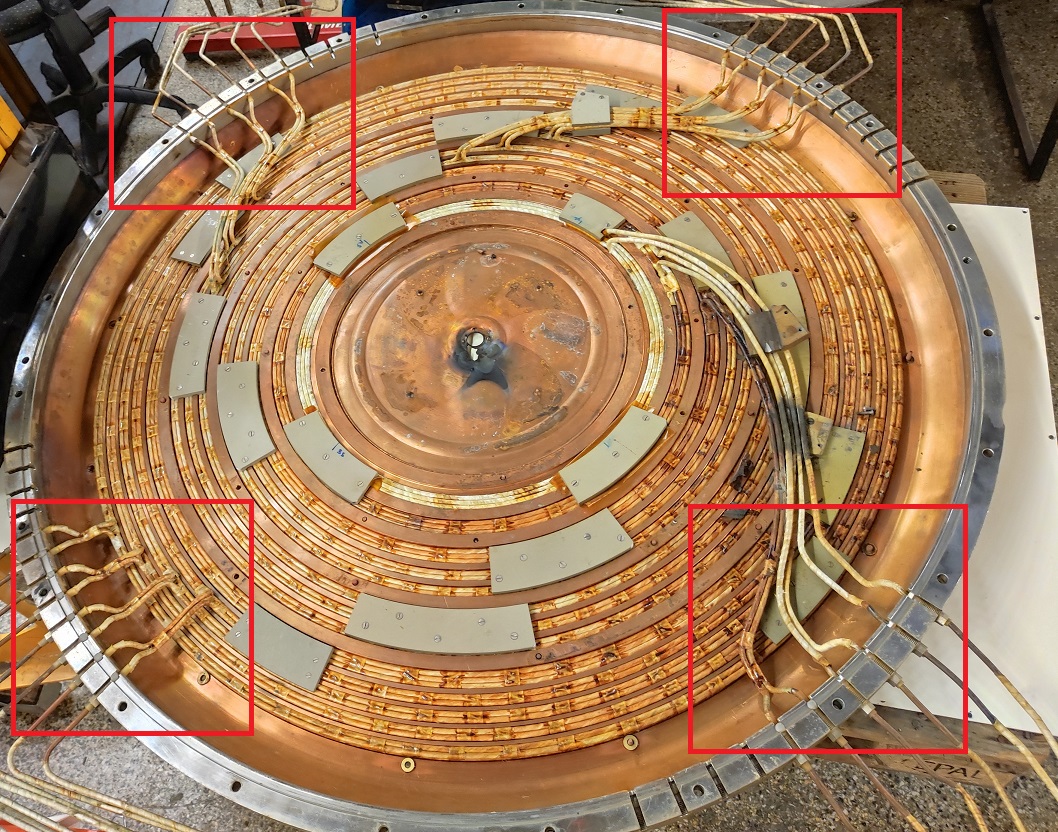}
\caption{trim-coils No. 5 -- 18 mounted in their copper holder, which serves also as a cover of the magnet sectors from Fig.\,\ref{fig:HC_coils_full}. The red squares indicate the location of the new harmonic coils. The dark spots are traces of the electrical breakdown of the trim-coil No.\,6 leads insulation. }
\label{fig:plakyr_hc}
\end{figure}

\begin{figure}[ht]
\centering
\includegraphics[width=0.47\textwidth]{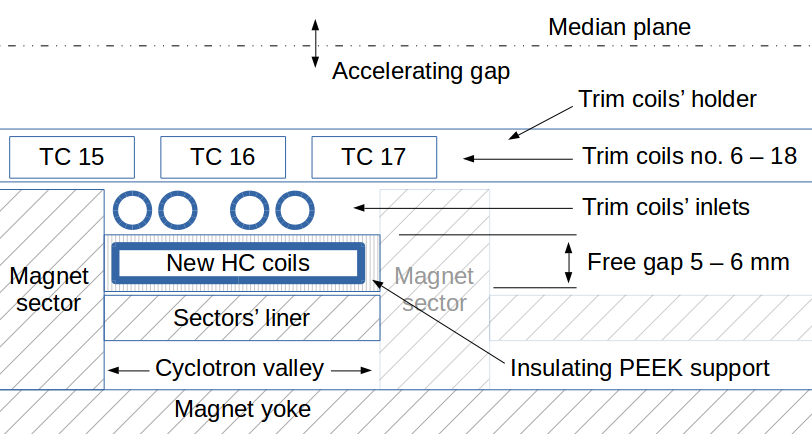}
\caption{Schematic plot of the installation gap for the new harmonic coils. Actual dimensions are not in scale.  }
\label{fig:hc_gap}
\end{figure}

The mechanical layout of the cyclotron magnetic system posed significant challenges for the design and installation of the new harmonic coils (HC). The U-120M magnetic structure is composed of iron spiral magnet sectors situated on magnet pole extensions and a set of 18 concentric trim-coils, which are installed inside the vacuum accelerating chamber, see Fig.\,\ref{fig:new_HC_in_chamber}. The first five trim-coils are integrated into the iron sectors and their radial positions reach up to 180\,mm. The remaining ones are part of the copper trim-coils' holder, as illustrated in Fig.\,\ref{fig:plakyr_hc}. Between the radii 480--560\,mm, the magnet sectors are horizontally connected by iron sector liners, see Fig.\,\ref{fig:hc_gap}, to shape the magnetic field at the highest radii. The new OHC were installed at the only viable location in the cyclotron valley space -- on top of the sector liners and beneath the trim-coils' holder, as schematically depicted in Fig.\,\ref{fig:hc_gap}. This space, marked by red squares in Fig.\,\ref{fig:plakyr_hc}, also accommodates the inlets for the trim-coils' current and water cooling copper tubes. It is important to note that these inlets are significantly age-hardened, and any mechanical intervention could risk micro ruptures, potentially leading to water leakage issues in the future. Additionally, any modification to the iron components of the cyclotron's magnetic structure, such as altering the sectors' liners, would have adverse effects on the resulting magnetic field. Hence, the only feasible solution was a non-invasive approach to install the coils within the existing magnetic system.

The available space for the OHC between the trim-coil inlets and sector liners was limited to 6\,mm at three OHC coil positions and to 5\,mm only at the fourth position. This limitation led to the design of two slightly different coil pairs, constructed from 4$\times$4\,mm and 5$\times$5\,mm hollow rectangular copper tubes. The remaining 1\,mm of free space is occupied by a plastic electrical insulating support made of PEEK and housing the coils. The arrangement of the new harmonic coils and the trim-coil power leads above the sector liners is illustrated in Fig.\,\ref{fig:hc_gap}.

\begin{figure}[ht]
    \centering
 \begin{subfigure}[b]{0.22\textwidth}
     \centering
    {\includegraphics[angle=90, height=7cm]{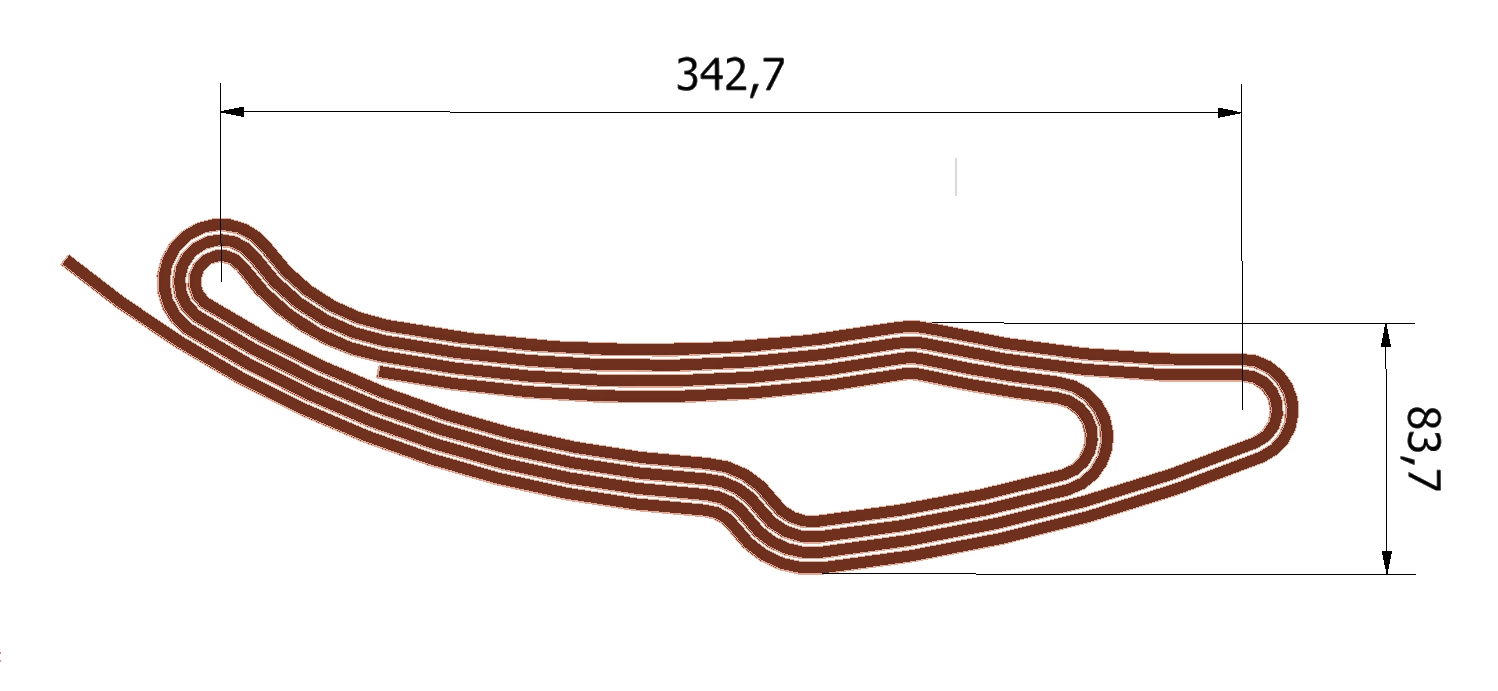}}
    \caption{4$\times$4~mm coil. }
    \label{subfig:hc1_dim}
\end{subfigure}
 \begin{subfigure}[b]{0.22\textwidth}
     \centering
    {\includegraphics[angle=90, height=7cm]{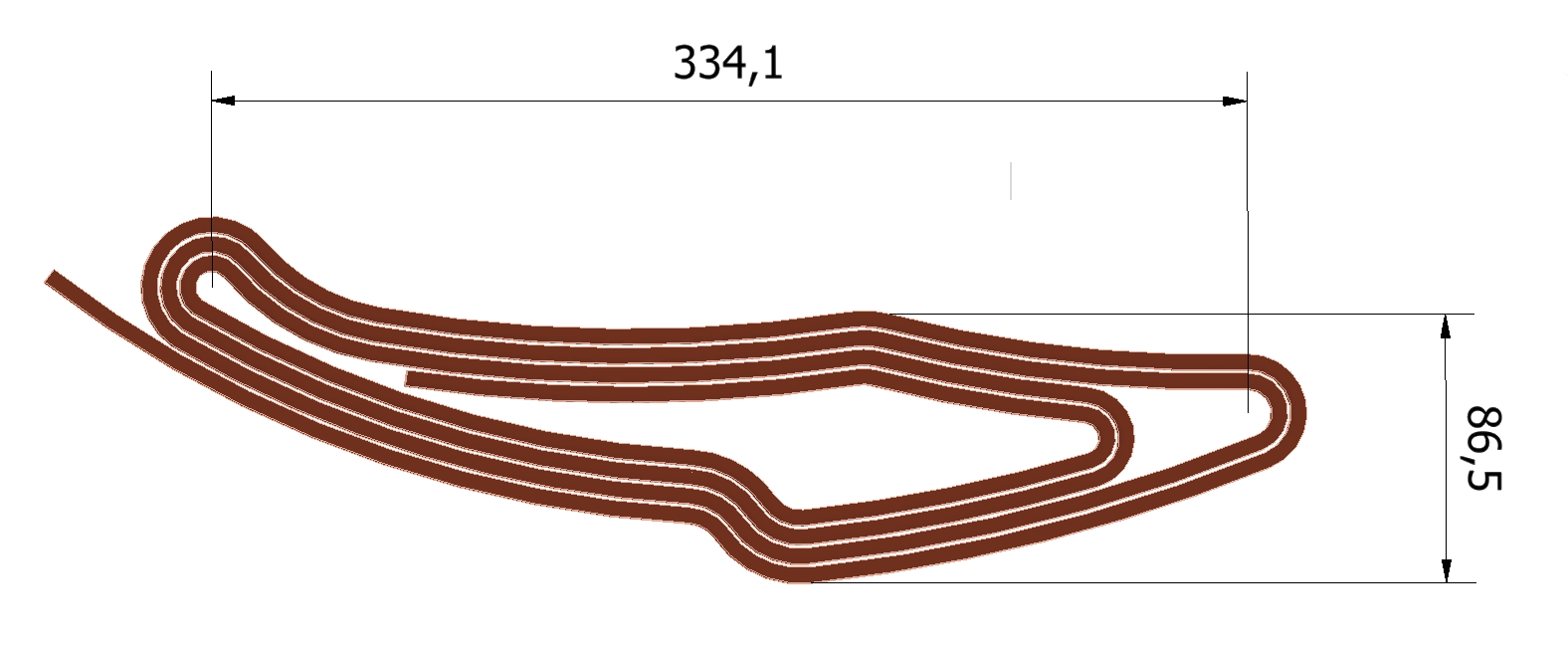}}
    \caption{5$\times$5~mm coil.}
    \label{subfig:hc2_dim}
 \end{subfigure}
    \caption{New harmonic coils. Dimensions are in mm.}
    \label{fig:hc_dim}
\end{figure}

To maximize the effect of the coils on the resulting magnetic field at the extraction radius, the coils are shaped to have the longest possible azimuthal length, occupying the maximum area between the magnetic sectors. Simultaneously it was necessary to suppress the influence of the coils on the beam at lower orbits by  minimizing their radial size. These two requirements were taken into account in the final shape of the coils, as shown in Fig.\,\ref{fig:hc_dim}.

\begin{figure}[ht]
    \centering
 \begin{subfigure}[b]{0.11\textwidth}
     \centering
    {\includegraphics[angle=90, height=5.6cm]{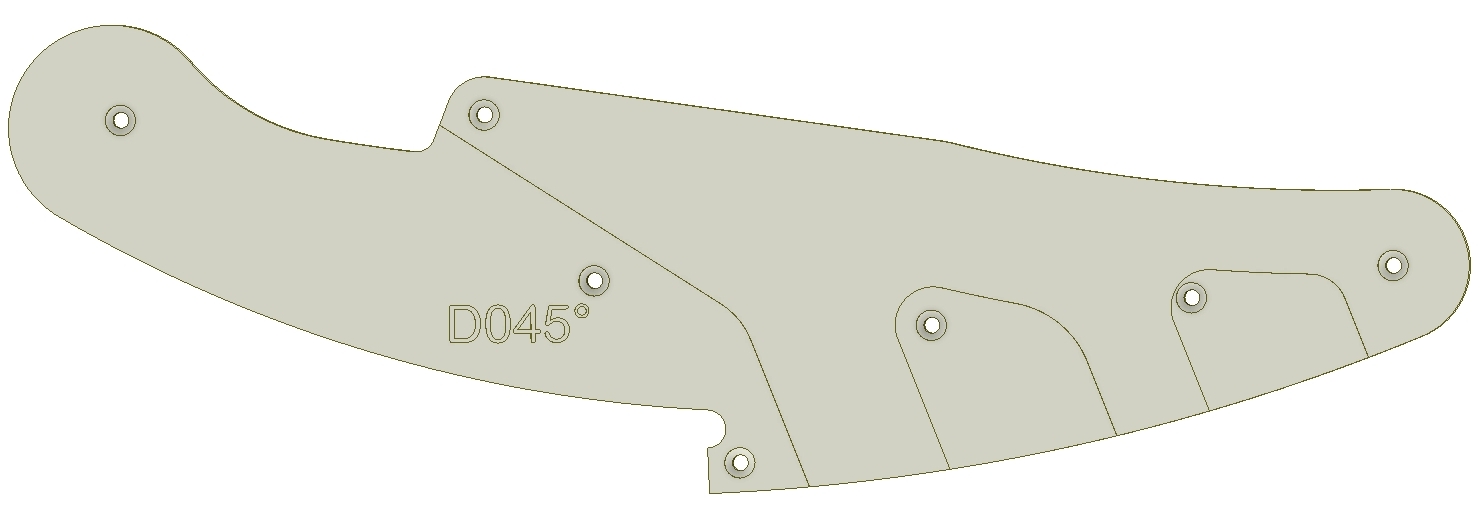}}
    \caption{45$^\circ$.}
\end{subfigure}
 \begin{subfigure}[b]{0.11\textwidth}
     \centering
    {\includegraphics[angle=90, height=5.6cm]{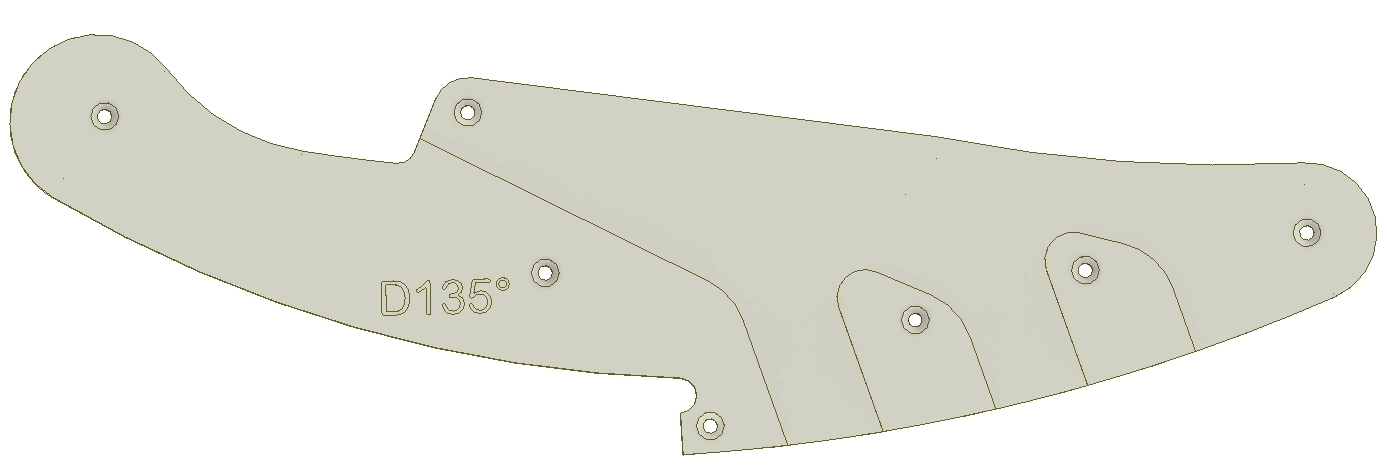}}
    \caption{135$^\circ$.}
 \end{subfigure}
  \begin{subfigure}[b]{0.11\textwidth}
     \centering
    {\includegraphics[angle=90, height=5.6cm]{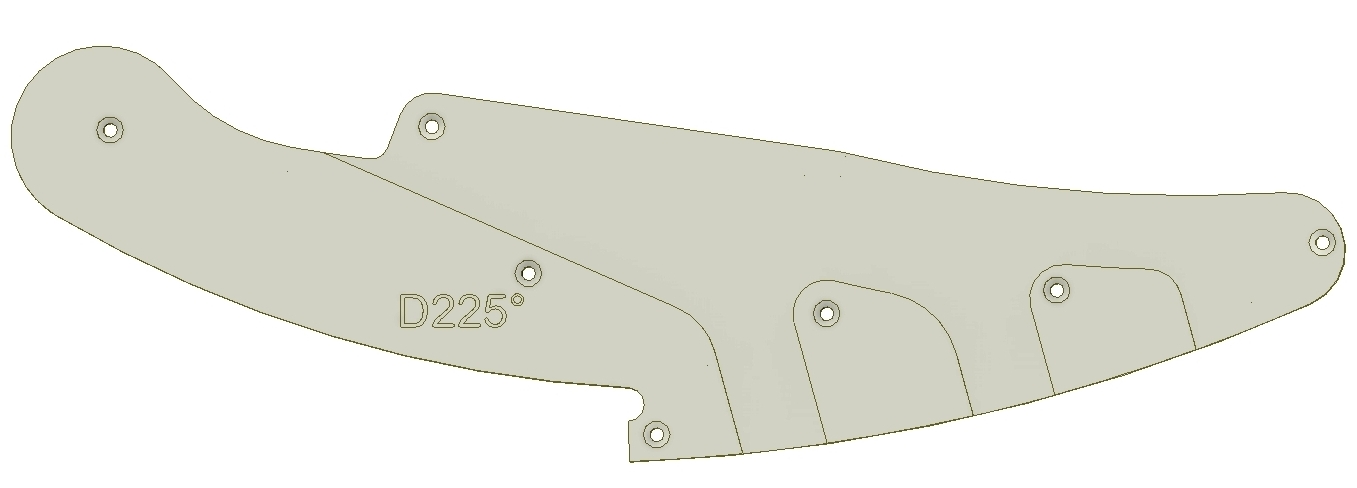}}
    \caption{225$^\circ$.}
\end{subfigure}
 \begin{subfigure}[b]{0.11\textwidth}
     \centering
    {\includegraphics[angle=90, height=5.6cm]{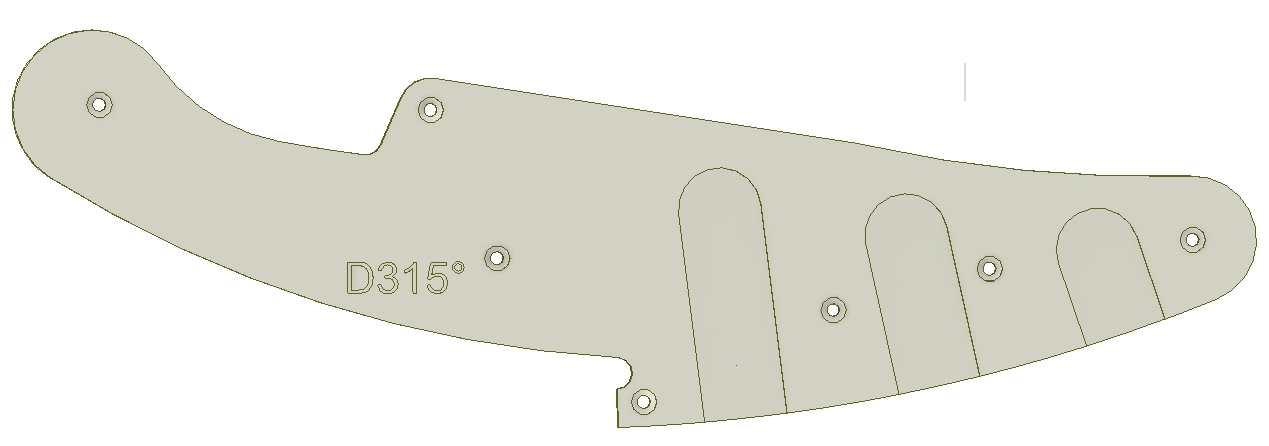}}
    \caption{315$^\circ$.}
 \end{subfigure}
    \caption{Design of the PEEK supports for bottom harmonic coils at azimuths of 45$^\circ$, 135$^\circ$, 225$^\circ$, and 315$^\circ$. Top view.}
    \label{fig:hc_supports}
\end{figure}

\begin{figure}[ht]
\centering
\includegraphics[width=0.47\textwidth]{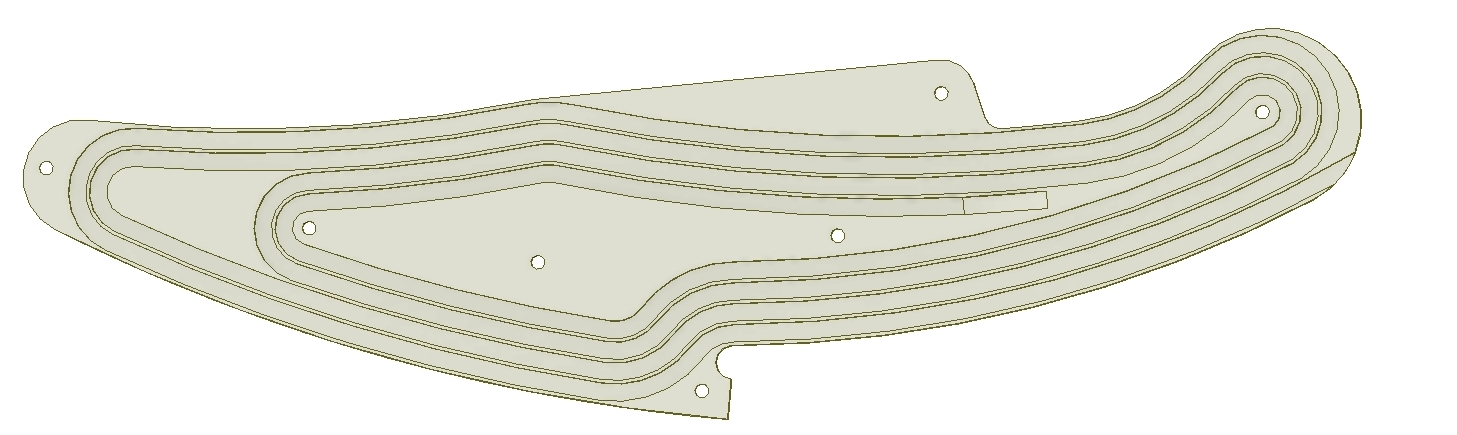}
\caption{Harmonic coil support at azimuth 225$^\circ$ with 1~mm thick wall for electric insulation between individual turns. Bottom view.  }
\label{fig:hc_support_bottom}
\end{figure}

\begin{figure}[ht]
\centering
\includegraphics[width=0.47\textwidth]{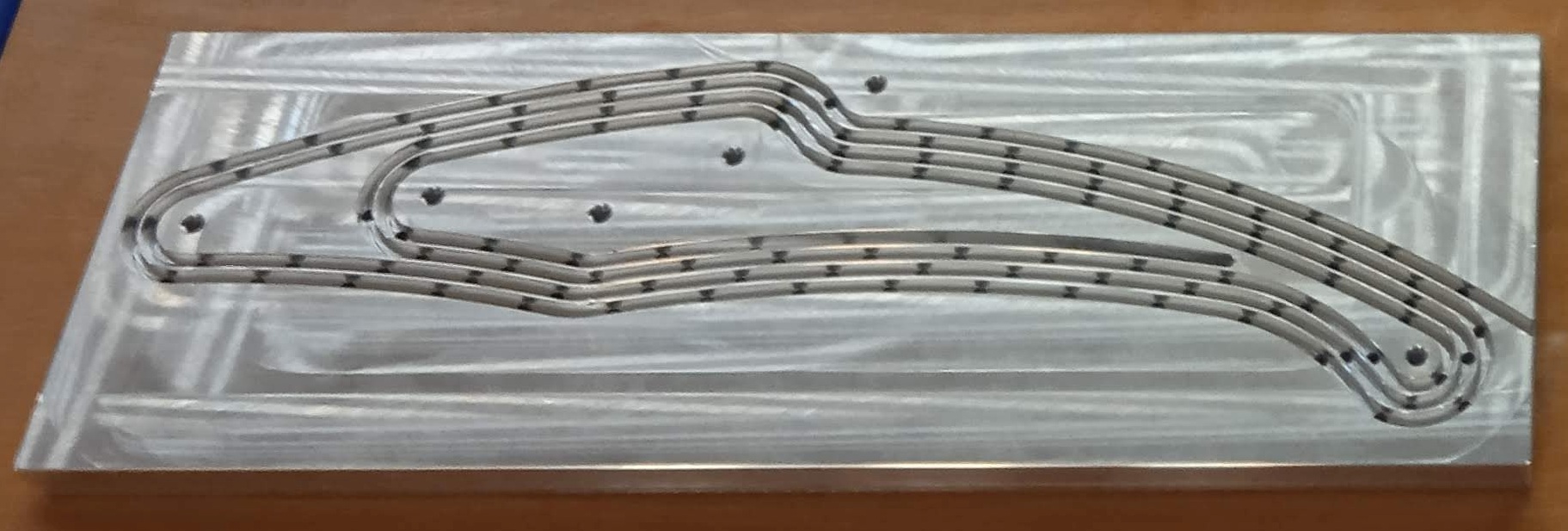}
\caption{Aluminum mold for forming the coils.  }
\label{fig:new_HC_formy}
\end{figure}

The constrained vertical clearances and the non-uniform arrangements of the trim-coils' leads, as depicted in Fig.\,\ref{fig:plakyr_hc}, necessitated the individual design of electrical insulation supports for each OHC coil as presented in Fig.\,\ref{fig:hc_supports}. The coils were shaped by pressing the rectangular copper tubes into an aluminum alloy mold shown in Fig.\,\ref{fig:new_HC_formy}. Afterwards they were transferred to their respective PEEK supports. These coils do not have additional electrical insulation, and the electrical insulation between the individual turns is provided by 1\,mm thick side and top walls created in the supports. The bottom side of the coils is insulated from the sectors' liners by a 0.25\,mm thick epoxy-fiberglass sheet. A bottom view of the support at 225$^\circ$ is depicted in Fig.\,\ref{fig:hc_support_bottom}.

\begin{figure}[ht]
    \centering
 \begin{subfigure}[b]{0.23\textwidth}
     \centering
    {\includegraphics[width=\textwidth]{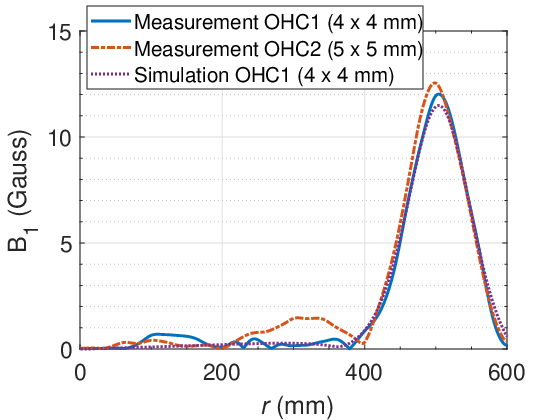}}
    \caption{Measured influence over radius and simulation. }
    \label{subfig:hc_compar_overal}
\end{subfigure}
 \begin{subfigure}[b]{0.23\textwidth}
     \centering
    {\includegraphics[width=\textwidth]{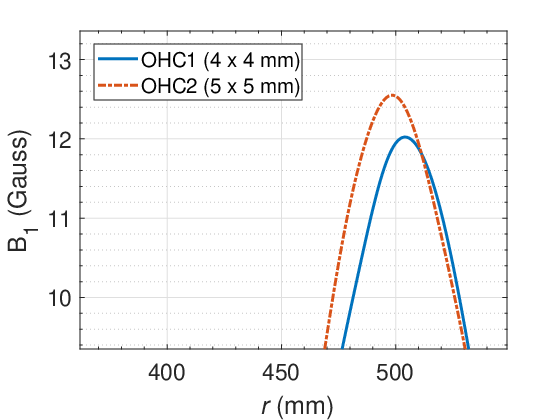}}
    \caption{Detail of the OHC coils' peak tip. }
    \label{subfig:hc_compar_detail}
 \end{subfigure}
    \caption{Comparison of the measured and simulated magnetic field first harmonic component of the new harmonic coils for current 200\,A. }
    \label{fig:hc_comparison}
\end{figure}

 The design of the harmonic coils configured as pairs of opposite coils with reversed polarity eliminates the average component of the magnetic field. To preserve the symmetry of the magnetic field, the harmonic coils are mirrored across the cyclotron median plane. Figure\,\ref{fig:hc_comparison} illustrates the first harmonic component of the  magnetic field produced by both pairs of coils, along with data simulated for the 4$\times$4\,mm coil (OHC$_1$). Simulation data of the 5$\times$5\,mm coil (OHC$_2$) is not shown, but it describes the measured data quantitatively similarly as in the case of OHC$_1$ coil. The simulations, performed in CST Studio \cite{CST}, are in good agreement with the measured shape of the resulting magnetic field.

The measured data also revealed the presence of a weak parasitic component between radii 200--400\,mm. This parasitic component is likely caused by an imperfect overlap of the supply-power lead tubes in the series connection of the OHC$_2$ coils, leading to the formation of a false coil. The parasitic contribution at radii 200--400\,mm amounts to approximately 15\% of the OHC$_2$ coil contribution. Due to the utilization of tubes with different cross-sections, the two pairs of harmonic coils exhibit a slightly different shape. The radial length of the coils differs by 3\,mm, and the coil's centers radial positions are offset by 4\,mm. In Fig.\,\ref{subfig:hc_compar_detail}, a detailed view of the peak tip at a radius of approximately 500\,mm is presented. The slight variation in the peak amplitude is attributed to the distinct geometry of both coils. The magnetic field map of the new harmonic coils is illustrated in Fig.\,\ref{fig:hc_maps}.

By setting the current amplitudes $I_{\mathrm{OHC}1}$ and $I_{\mathrm{OHC}2}$, it is possible to induce a perturbation in the magnetic field \cite{Joho, schwabe}, denoted as 

\begin{equation}
    \Delta B_N(r,\theta) = b_N(r)\cos(\theta-\theta_N(r)),
    \label{eq:b_coil}
\end{equation}

\noindent where $b_N(r)$ and $\theta_N(r)$ is the amplitude and the phase of the $N$-th harmonic of the magnetic field at the radius $r$. The first harmonic component produced by the harmonic coils $\Delta B_{1\mathrm{OHC}}$ is represented by an amplitude  $b_{1\mathrm{OHC}} = k\sqrt{I_{\mathrm{OHC}1}^2 + I_{\mathrm{OHC}2}^2}$, and an azimuth $\theta_{1\mathrm{OHC}} = \arctan( I_{\mathrm{OHC}1}/I_{\mathrm{OHC}2})$. For the fully saturated cyclotron magnet, the coefficient $k$ is approximately 0.06\,Gauss/A.



\section{Modification of the deflecting system}
\label{sec:extraction}

The original positive-ion extraction system was designed as a modification of the regenerative beam extraction concept mostly employed in synchro\-cy\-clo\-trons \cite{Joho, victor-cyklo}. This modification involves the suppression of the second harmonic component of the magnetic field, ensuring the conservation of radial beam emittance during the extraction while maintaining an extraction efficiency over 60\% \cite{trejbal4-system_vozbuzdenja}. The extraction concept was specifically developed for the U-120M cyclotron, the first isochronous cyclotron in the Eastern block \cite{trejbal3-rascet_elektrostat}. The system was fully implemented in an electrostatic configuration, consisting of three electrostatic deflectors positioned at azimuths $210^\circ$, $272^\circ$, and $305^\circ$, along with an electrostatic exciter and its compensator. The electrostatic exciter (peeler), responsible for generating coherent radial oscillations of the beam and the associated outward radial impulse, was positioned at azimuth $182^\circ$, just before the first deflector. The second harmonic compensator was placed at $90^\circ$ w.r.t. the exciter, serving as a part of the second deflector. All deflectors in the system were constructed with profiled high-voltage electrodes to ensure radial focusing of the beam passing through the cyclotron's fringe field. 

In the 1980s, due to uncontrollable sparking, the original electrostatic exciter was replaced with a bump-coil. Shortly thereafter, an accidental melting of the compensator occurred, leading to the operation of the extraction system as a single bump-coil system. This change resulted in a significant drop in extraction efficiency, reducing it to the actual value of approximately 10--15\%.


\subsection{New extraction system concept}

\begin{figure}
\centering
\includegraphics[width=0.487\textwidth]{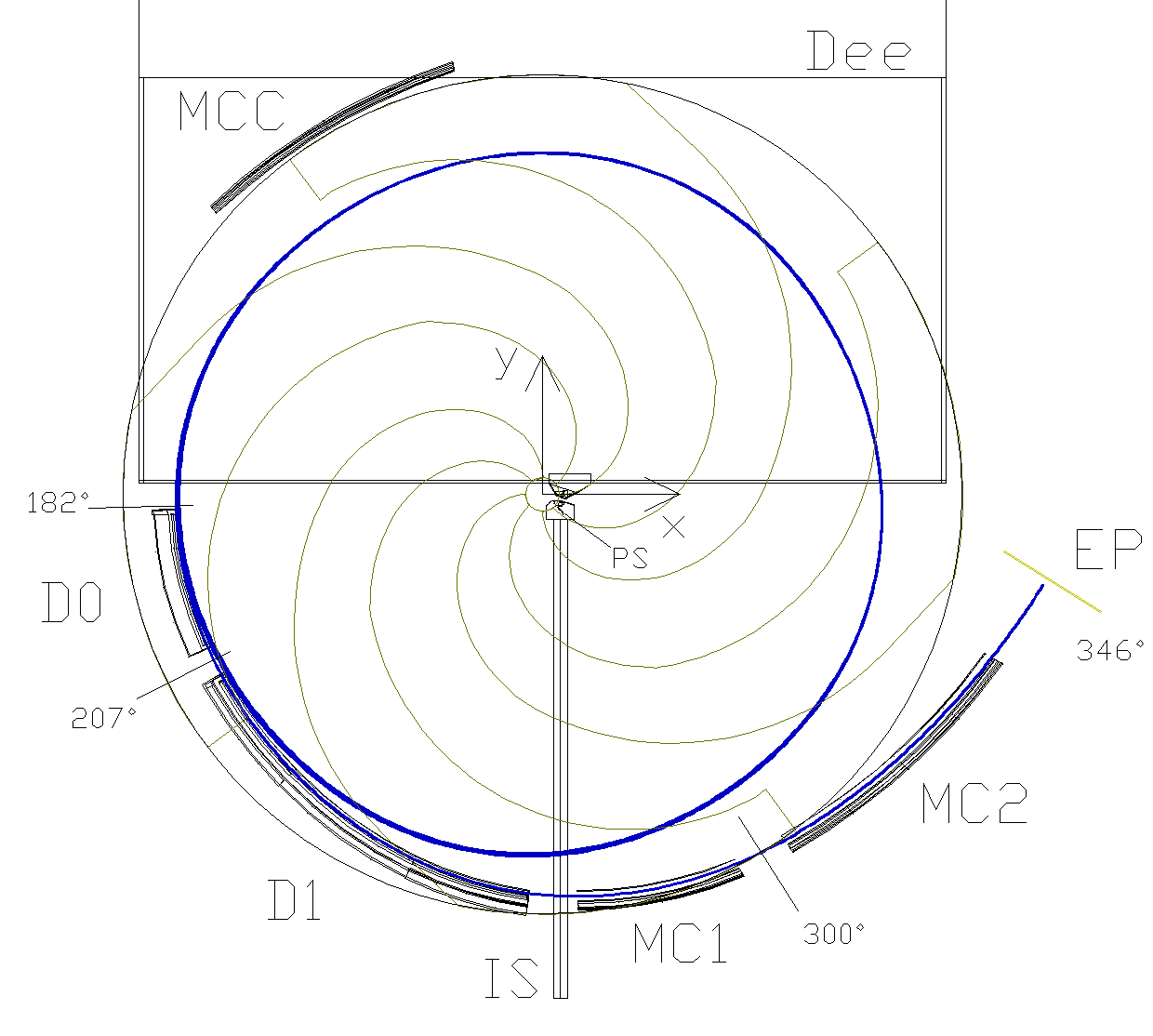}
\caption{Layout of the new extraction system. D0, D1 -- electrostatic deflectors, MC1, MC2 -- magnetic channels, MCC -- magnetic channels compensator, EP -- extraction point. Phase slits PS acting at the first ions orbit are part of the ion-source IS in the cyclotron center. The planes for investigating the beam properties between the extraction elements are marked with their azimuths.  }
\label{fig:new_ext}
\end{figure}

 The concept of the new extraction system comprises two electrostatic deflectors and two magnetic channels, as illustrated schematically in Fig.\,\ref{fig:new_ext}. The new deflector, referred to as D1, is positioned at the same location as the old first deflector, and its azimuthal length was also preserved.

The space previously occupied by the exciter has been filled by a new short deflector, designated as D0, with an azimuthal length of 22$^\circ$. A similar extraction system concept is employed at the AIC--144 cyclotron in Cracow, where it operates reliably \cite{krakow2}. The role of the electrostatic exciter, which was to generate coherent radial oscillations and provide orbit separation at the entry to the first deflector, has been taken over by the newly installed harmonic coils.

In the initial stage of designing the new extraction system concept, two primary extraction methods -- the \emph{integer resonance method} for $\nu_r=1$ and the \emph{precessional extraction} were investigated \cite{Joho, Heikkinen, Hagedoorn2, schwabe}. These methods differ from non-linear half-integer regenerative extraction methods primarily by the absence of an element for altering the second harmonic magnetic field component \cite{trejbal4-system_vozbuzdenja}. 

The integer resonance method involves extracting the beam either before it enters the $\nu_r=1$ resonance or precisely at this resonance point. In the literature, it is sometimes referred to as a ''brute force'' or a linear resonance extraction method \cite{schwabe}. This method  enables the extraction of a broad RF phase beam, resulting in high overall extraction efficiency while  preserving the emittance during the extraction process. However, this method comes at the cost of requiring a significantly higher magnetic field bump and, since the beam is extracted from a region where $\nu_r \geq 1$, the extraction radius and thus the energy of the extracted beam are reduced, necessitating the maintenance of higher electric fields in the deflectors.

The precessional extraction method relies on accelerating the beam to the fringe field region where $\nu_r < 1$, utilizing a very subtle first harmonic bump. This method is well-established and widely used in most modern cyclotron facilities. The pre\-ce\-ssion me\-thod provides the maximum possible final beam energy for a given cyclotron with reasonable extraction efficiency, often enhanced by restricting the RF phase size using phase slits. In this method, the extraction radius is higher compared to the integer resonance method, and the required voltage on deflectors can be lower. A general comparison of the two methods is discussed in Ref. \cite{liem}.

\subsection{Numerical simulation of the extraction process}

The numerical simulations of the acceleration and extraction processes are carried out using two distinct tools: the in-house developed beam dynamics calculator DuryCNM \cite{cihak_dur}, specially tailored for simulating the U-120M cyclotron, and the universal beam dynamics analysis code for compact cyclotrons, SNOP \cite{snop}. Both calculators employ Runge-Kutta method-based solvers for the equations of ion motion in an electromagnetic field.

The SNOP code allows for the import of electromagnetic fields of the individual accelerator elements, such as the dee, harmonic coils, electrostatic deflectors, and magnetic channels together with their 3D CAD models, enabling a precise evaluation of particle losses during the acceleration and extraction processes. To facilitate this, CAD models of the ion-source with the puller, phase slits, accelerating electrode, and individual extraction elements were created in the Inventor software \cite{inventor} and analyzed using the 3DS SIMULIA electromagnetic simulators Opera and CST Studio \cite{CST}. The imported electromagnetic fields of the extraction elements encompass not only the functional parts of the electromagnetic fields involved directly in the extraction process but also their fringe fields, which strongly affect the accelerated beam.


The magnetic fields used to simulate beam dynamics were taken from a complex mathematical model of the U-120M cyclotron Wmodel \cite{ cihak2} that is based on precise magnetic measurements conducted in the 1980s. The model calculates the required settings of the magnet and trim-coils' currents related to the isochronous field for a specific particle type and chosen final energy.

The simulations are performed for a single particle with a specific RF phase, position, and momentum or for a particle bunch composed of 10,000 ions. The bunch is defined within the full cyclotron RF phase interval of $\pm 90^\circ$ with initial energy ranging from 10--50\,eV. The radial and vertical emittances $\varepsilon_r$ and $\varepsilon_z$ are set to 180 and 100\,mm\,mrad, respectively \cite{forringer_tppe012}. Particles colliding with the central region structures (puller, ion-source, and phase slits) are removed within the first several turns, and the remaining particles form an internal cyclotron bunch, as illustrated in Fig.\,\ref{subfig:preces_bar}. The simulations start the acceleration at the ion-source slit and the bunch is then accelerated to the extraction radius with the dee voltage tuned in a range 34--35\,kV with preset harmonic-coil currents and scaled electromagnetic fields of the extraction elements. After passing the extraction system, the beam is directed to the cyclotron extraction point ($EP$). Parameters of the beam such as vertical emittances and the cross section in the  $r-z$ plane are evaluated at the $EP$ as well as on planes between the individual extraction elements (see Fig.\,\ref{fig:new_ext}).

The extraction efficiency $\eta$ is calculated as the ratio of the number of particles, which are extracted from given bunch ($N_{\rm{extracted}}$), to the number of particles that were in the internal bunch when the beam is constrained by the phase slits ($N_{\rm{internal}}$)

\begin{equation}
    \eta = \frac{N_{\rm{extracted}}}{N_{\rm{internal}}}.
    \label{eq:eff_part}
\end{equation}

In addition, one can define the overall extraction efficiency 

\begin{equation}
    \eta_{\rm{T}} = \frac{N_{\rm{extracted}}}{N_{\rm{T}}},
    \label{eq:eff_tot}
\end{equation}

\noindent which is calculated as the ratio of the extracted part of the bunch $N_{\rm{extracted}}$ to the number of particles in the internal bunch $N_{\rm{T}}$, when  the phase slits are not used, i.e. the
maximal possible internal bunch for given cyclotron
central region geometry.

The radial size of the accelerated bunch is minimized by appropriate centering the particle with RF phase 0$^\circ$, achieved through the use of two pairs of internal harmonic coils (IHC), see Fig.\,\ref{fig:HC_coils_full}. The centering is optimized for each particle type and extraction mode, ensuring that all extraction efficiency simulations are calculated for the minimal radial size of the beam.


\section{Extraction system optimization}
\label{sec:results}

\begin{figure}
\centering
\includegraphics[width=0.48\textwidth]{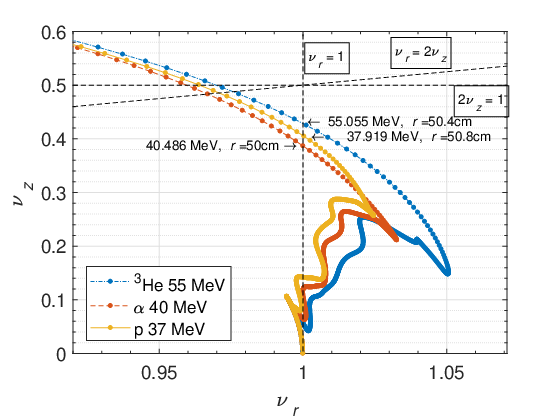}
\caption{Tune diagrams for the maximum energy and three ion modes of the U-120M cyclotron.}
\label{fig:tunes}
\end{figure}

The optimization of the system and the estimation of extraction efficiency were performed for the 37\,MeV proton mode and for the 55\,MeV helium-3, mode which correspond to the maximum output energies of the cyclotron. The corresponding tune diagrams, with the highlighted radii, where $\nu_r$ crosses unity, are shown in Fig.\,\ref{fig:tunes}. As the internal beam intensities do not exceed a few tens of microamperes for helium and a hundred microamperes for protons, no space-charge effect is considered in the calculations.

\subsection{Orbit separation}

The amplitude and the phase of the first harmonic magnetic field component generated by the outer harmonic coils ($\Delta B_{1\mathrm{OHC}}$) are given by the current amplitudes $I_{\mathrm{OHC}1}$ and $I_{\mathrm{OHC}2}$, see Eq. \ref{eq:b_coil}. These currents were tuned manually for optimal orbit separation azimuth leading to maximum  extraction efficiency in both types of extraction. For the precessional extraction ($PE$), the additional first harmonic component generated by the harmonic coils falls in the range of 3--5\,Gauss. In contrast, for linear resonance extraction ($LE$), this component is substantially higher, approaching 50\,Gauss.

\begin{figure}[t]
    \centering
\begin{subfigure}[b]{0.23\textwidth}
     \centering
    {\includegraphics[width=\textwidth]{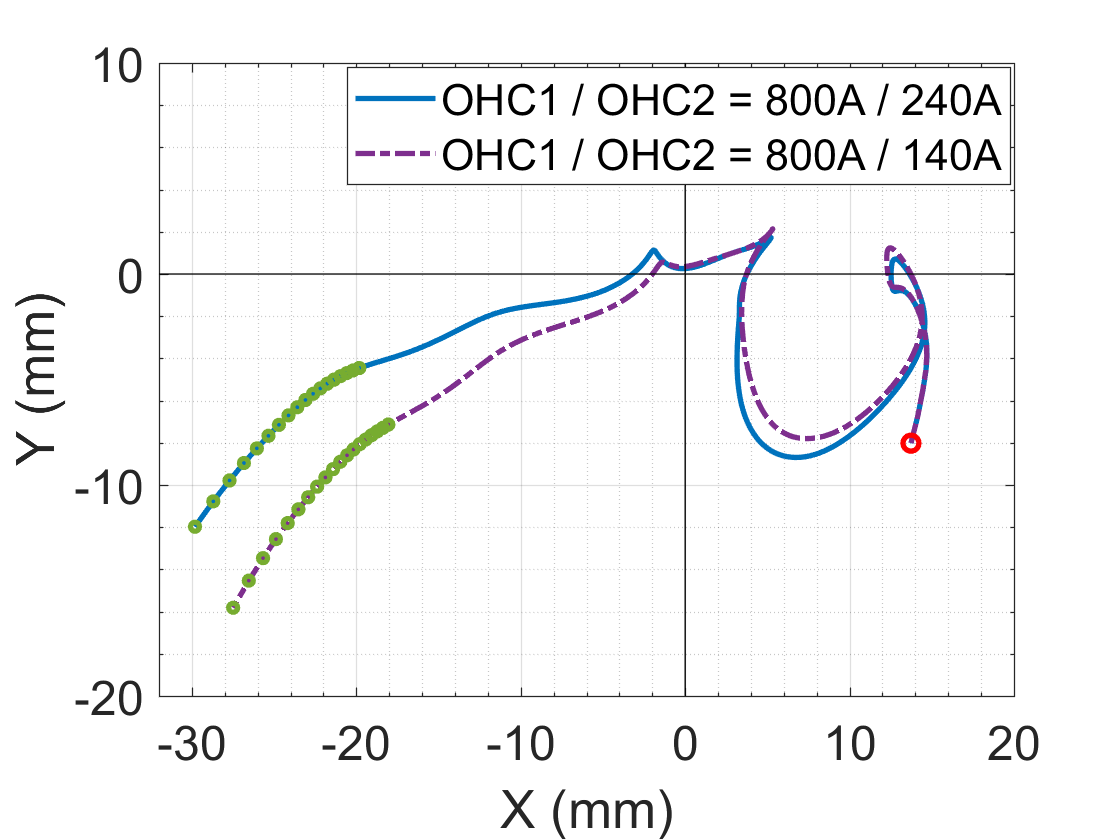}}
    \caption{Linear resonance extraction.}
    \label{subfig:center_le}
\end{subfigure}
    \hfill
 \begin{subfigure}[b]{0.23\textwidth}
     \centering
    {\includegraphics[width=\textwidth]{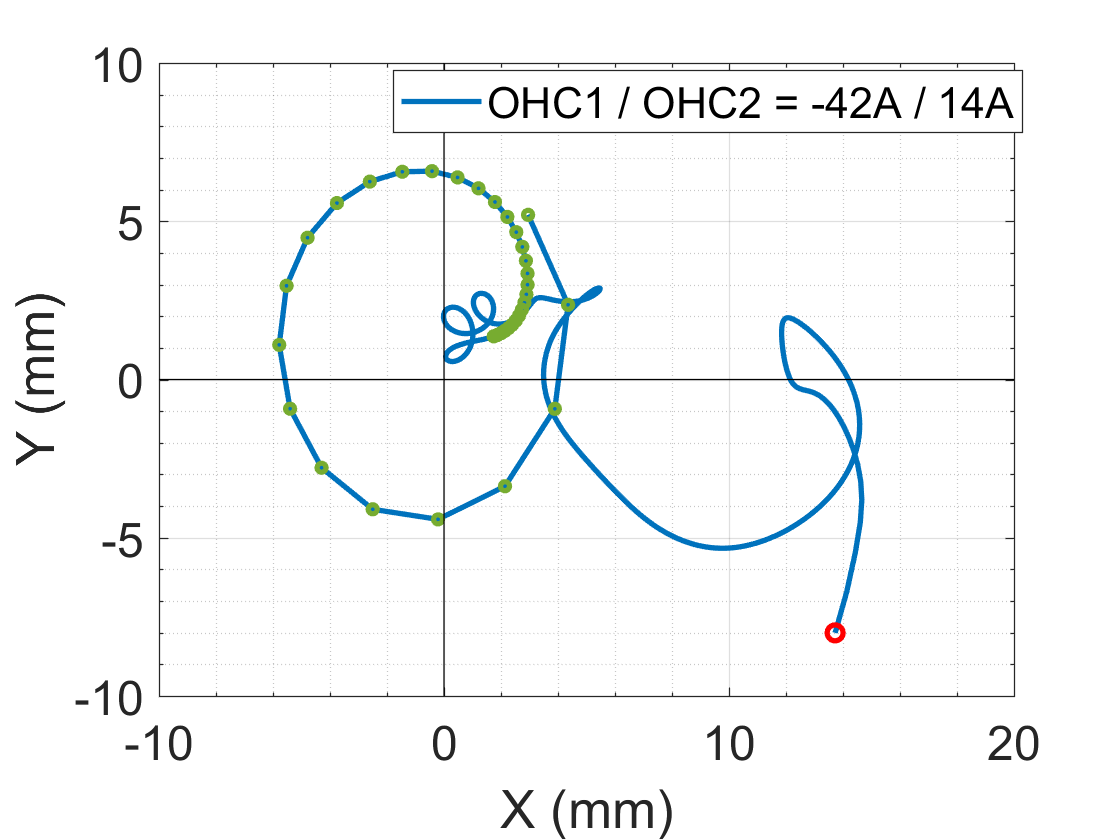}}
    \caption{Precessional extraction.}
    \label{subfig:center_pe}
\end{subfigure}

\begin{subfigure}[b]{0.236\textwidth}
     \centering
    {\includegraphics[width=\textwidth]{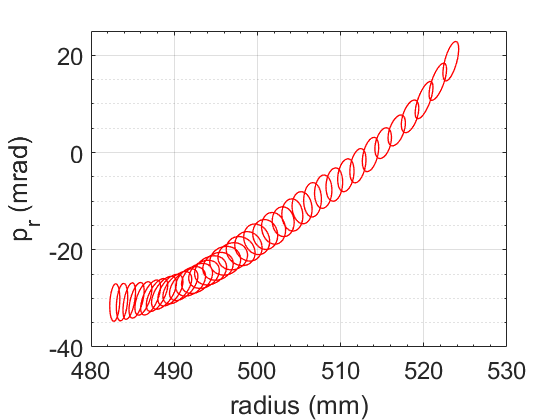}}
    \caption{Linear resonance extraction.}
    \label{subfig:pr_res_free}
\end{subfigure}
    \hfill
 \begin{subfigure}[b]{0.236\textwidth}
     \centering
    {\includegraphics[width=\textwidth]{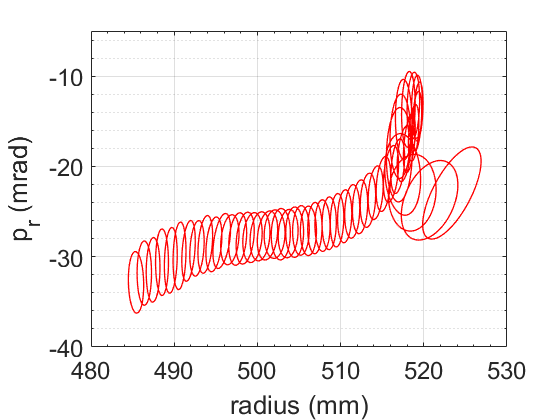}}
    \caption{Precessional extraction.}
    \label{subfig:pr_precess}
\end{subfigure}

\begin{subfigure}[b]{0.236\textwidth}
     \centering
    {\includegraphics[width=\textwidth]{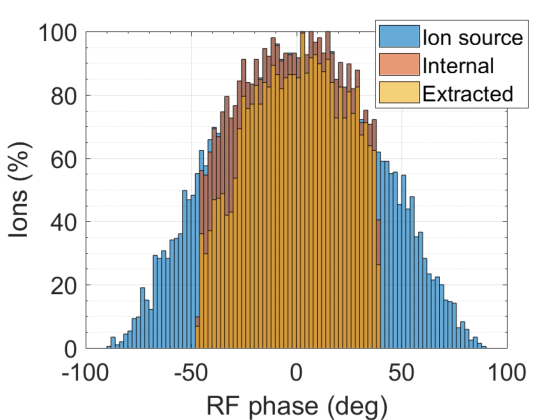}}
    \caption{Linear resonance extraction.}
    \label{subfig:res_free_bar}
\end{subfigure}
    \hfill
 \begin{subfigure}[b]{0.236\textwidth}
     \centering
    {\includegraphics[width=\textwidth]{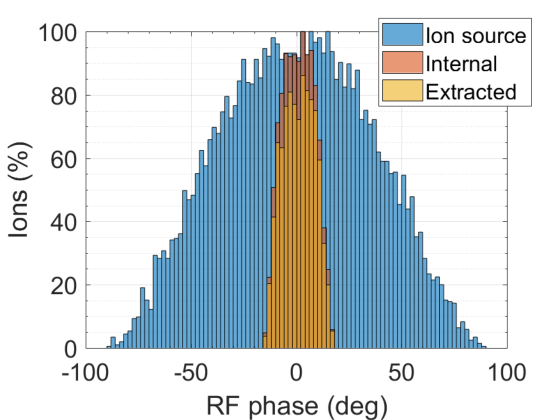}}
    \caption{Precessional extraction.}
    \label{subfig:preces_bar}
\end{subfigure}
    \caption{Comparison of the linear resonance ($LE$) and the precessional extraction method ($PE$) in the $^3$He 55\,MeV cyclotron mode:  \\  
    (a,b) -- Position of the orbital center of a RF phase central particle. The first orbit is marked by the red circle. \\
    (c,d) -- The $r-p_r$ phase space ellipse evolution for a bunch with the RF phase size 5$^\circ$. \\
    (e,f) -- Distribution of particles which form the extracted bunch within the RF space of the internal bunch.\\
    For the 37\,MeV proton mode, the figures are analogous.}
    \label{fig:centers_positive}
\end{figure}

In the case of $LE$, the degree of orbit separation depends on the amplitude of the $\Delta B_{1\mathrm{OHC}}$ component and its azimuth. Through the optimization of the $\Delta B_{1\mathrm{OHC}}$ amplitude and phase, it is possible to control drift direction of orbit centers and find the azimuth at which the turn separation for a given orbit is maximal \cite{Joho, schwabe}. Since the first deflector is located at 182$^\circ$, we investigated two possible center drift directions, as it is shown in Fig. \ref{subfig:center_le}: one in the direction of approximately 200$^\circ$ marked as $LE_1$, and the second in the direction of approximately 240$^\circ$ marked as $LE_2$. The $LE_1$ settings provides the largest separation of the orbits at the position of the first deflector D0. The advantage of the $LE_2$ setup is that it requires a much lower electric field in the main D1 deflector.  However, the $LE_2$ setup also has a lower extraction efficiency because it leads to a lower separation of the orbits at the extraction radius. The harmonic coil currents $I_{\mathrm{OHC}1}$ and $I_{\mathrm{OHC}2}$  for the three extraction modes are listed in Table \ref{tab:ext_params}.

The orbital center evolution for the $PE$, $LE_1$, and $LE_2$ extraction modes is shown in Fig.\,\ref{fig:centers_positive}. The harmonic-coil currents and the corresponding generated $\Delta B_{1\mathrm{OHC}}$ component are provided in Table \ref{tab:hc_params}. The table also shows the basic characteristics of the beam for the investigated extraction methods, along with the positions of the septum of the first deflector D0 and the electric fields of deflector D1. The values reported in Table \ref{tab:hc_params} correspond to the 37\,MeV  proton mode and the 55\,MeV helium-3 mode. The extraction efficiency is discussed in the next section.

\begin{table*}
\centering
\caption{Extraction system settings and extracted beam properties for the proton mode 37\,MeV. Values for the helium-3 mode 55\,MeV are in brackets where appropriate. }
\begin{tabular}{@{}llll@{}}\toprule
Extraction mode & $PE$ & $LE_1$ & $LE_2$  \\
\hline
    OHC currents $I_{\mathrm{OHC}1}$ / $I_{\mathrm{OHC}2}$ (A)  &  $-10$/50 ($-42$/14) & 800/240 & 800/80 (800/140)\\
    $\Delta B_{1\mathrm{OHC}}$ amplitude / phase (Gauss / deg) &  3.2/156 (2.7/214) & 50/70  & 48/59 (48/63) \\
    Radial beam size (mm) & 3.1 (5.2)  & 2.1 (2.9) & 1.8 (2.7)  \\
 
    Turn separation at 182$^\circ$(mm) &2.8 (4.4) & 1.5 (2.6) \ & 1.3 (2.3) \\
    Deflector D0 septum radius (mm) &  532 (527) & 524  & 524 \\
    Deflector D0 electric field (kV/cm) &  95 (78) & 125  & 125 \\
    Deflector D1 center electric field (kV/cm) &  95 (78) & 150 (113)& 122 (98) \\
\hline    
    Output energy (MeV) &  40.4 (58.9)& 36.5 (51.9) & 36.8 (52.1) \\
    Output energy FWHM (MeV) &0.06 (0.13) &0.24 (0.41)& 0.24 (0.41)\\
    Extraction efficiency  $\eta$  &  $>$\,70\% ($>$\,80\%) & $\sim$\,80\% & $\sim$\,70\% \\
\bottomrule
\end{tabular}
\label{tab:hc_params}
\end{table*}


\subsection{Extraction elements properties}

During the design of the new extraction system we took into consideration the need to maintain the operating parameters of the deflectors and magnetic channels within feasible limits. As the highest electric field intensities in the deflectors are required for extracting protons with the maximum energy, the design of the extraction system was optimized for the 37\,MeV proton $LE_2$ extraction mode. The other accelerating modes have lower demands on the electric field of deflectors.

\begin{figure}
    \centering
\begin{subfigure}[b]{0.22\textwidth}
     \centering
    {\includegraphics[width=\textwidth]{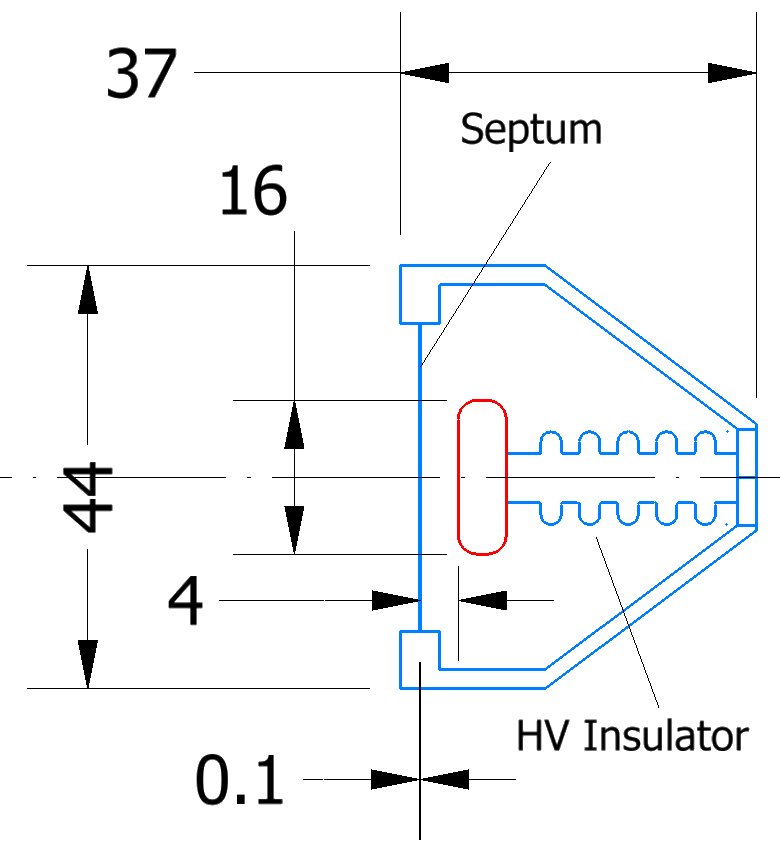}}
    \caption{Deflector D0}
    \label{subfig:d0_profile}
\end{subfigure}
    \hfill
 \begin{subfigure}[b]{0.25\textwidth}
     \centering
    {\includegraphics[width=\textwidth]{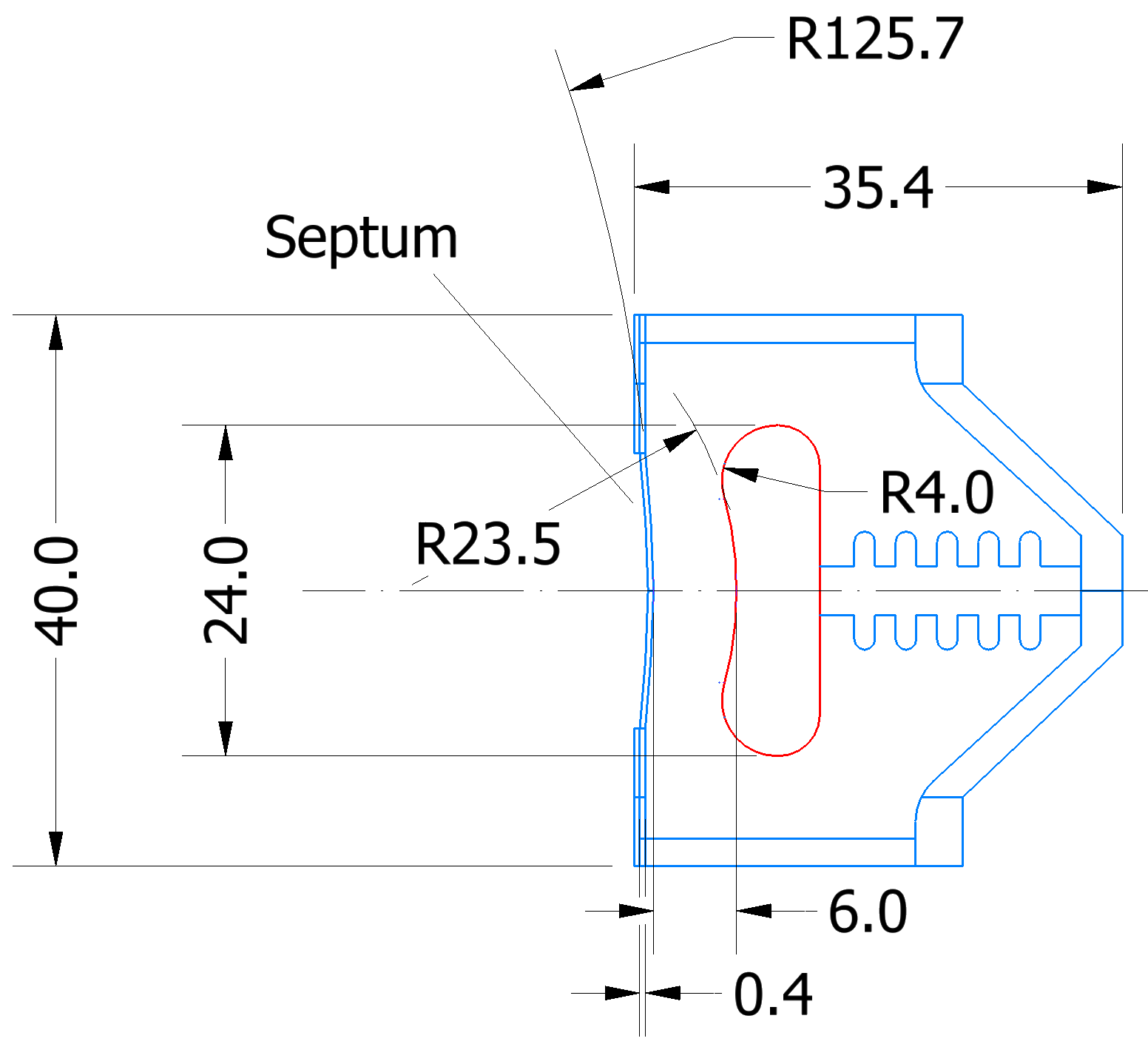}}
    \caption{Deflector D1}
    \label{subfig:d1_profile}
\end{subfigure}
    \caption{Radial cross section of the deflector D0 and D1. All dimensions are in mm. The high-voltage electrodes have red color,  the cyclotron median plane is represented by the dashed horizontal line. }
    \label{fig:d_profiles}
\end{figure}

The radial separation of the electrodes in the deflector D0 is chosen to be 4\,mm, which is large enough to accommodate the entire radial width of the proton beam in the $PE$ and the $LE$ extraction modes, ensuring a small margin. According to Botman and Hagedoorn \cite{Hagedoorn2}, the voltage limit for the 4\,mm gap size is approximately 70\,kV. For the proton mode 37\,MeV, the highest necessary D0 voltage is 55\,kV, well below the expected limit. The deflector D0 shape is adapted to the characteristics of the beam at the extraction radius in a way that losses are present only at the deflector's septum, and no beam is in contact with the high-voltage electrode. The septum is designed as a flat electrode with a thickness of 0.1\,mm. The material used for the septum will  be tungsten due to its high melting point. The deflector D0 high-voltage electrode has a round-corner rectangular cross-section profile, as seen in Fig.\,\ref{subfig:d0_profile}.

\begin{figure}[ht]
\centering
\includegraphics[width=0.36\textwidth]{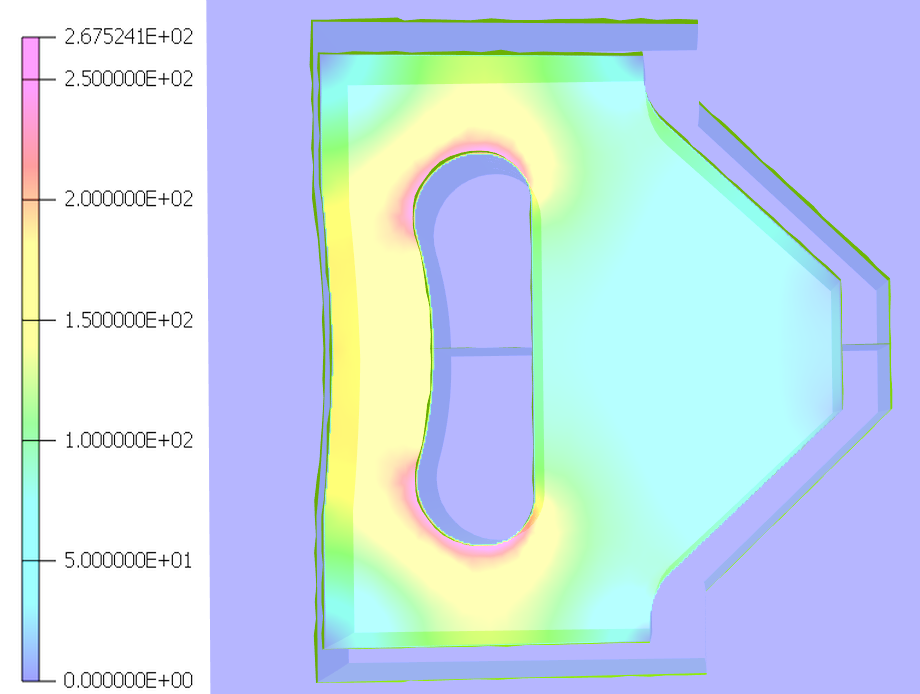}
\caption{Electric field distribution inside the new radially focusing deflector D1. The color scale is in V/m, HV electrode potential is +1\,V. }
\label{fig:d1_524_efield_map}
\end{figure}

\begin{figure}[ht]
    \centering
\begin{subfigure}[b]{0.235\textwidth}
     \centering
    {\includegraphics[width=\textwidth]{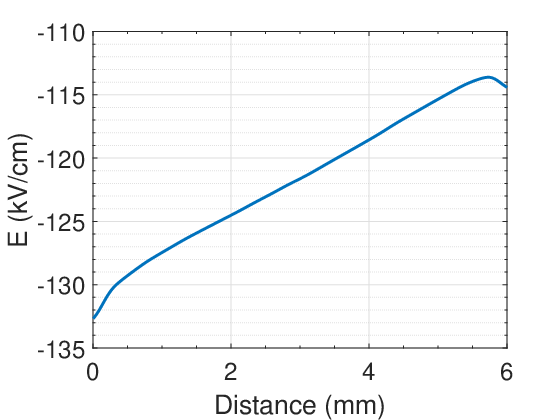}}
    \caption{Electric field.}
    \label{subfig:d1_e}
\end{subfigure}
    \hfill
 \begin{subfigure}[b]{0.235\textwidth}
     \centering
    {\includegraphics[width=\textwidth]{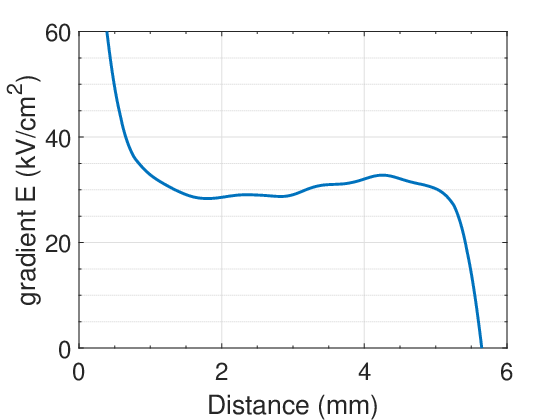}}
    \caption{Electric field gradient.}
    \label{subfig:d1_524_grad}
\end{subfigure}
    \caption{Deflector D1 electric field and its gradient with $-73$\,kV at the HV electrode for the 37 MeV proton mode. The horizontal axis is the distance from the septum electrode.}
    \label{fig:d1_524_efield}
\end{figure}

Already in the initial stages of the design, it became clear that for effective beam extraction, it is necessary to preserve the focusing properties of the deflector D1 and thus limit the radial width of the beam entering the subsequent section with the magnetic channels. Considering the Smith and Grunder criterion \cite{Grunder}, the $VE$ values of the new deflectors are set well below the ${VE}$ limit of $1.5\times10^{-4} \mathrm{kV^2/cm}$. For both calculated deflectors, the $VE$ value is approximately $1\times10^4\mathrm{kV}^2$/cm. As in the case of cyclotrons in other laboratories, the deflectors are designed enclosed to maintain the ability of high-voltage training before operation \cite{k1200}. Due to non-trivial beam trajectory inside the deflector D1, which is partially placed in the cyclotron valley, see Fig.\,\ref{fig:new_ext}, the deflector D1 is divided into three azimuthal sections of approximately equal length, each with a slightly different bending radius. By proper shaping the deflector D1 septum and the high-voltage electrode profile, it was possible to optimize the electric field gradient to be almost constant. The distribution of the electric field for a voltage of +1\,V  at the HV electrode is shown in Fig.\,\ref{fig:d1_524_efield_map}. The electric field and its gradient for a voltage of $-73$\,kV are shown in Fig.\,\ref{fig:d1_524_efield}.

Basic parameters of the new deflectors and magnetic channels optimized for 37\,MeV protons and the $LE_2$ extraction mode are listed in Table \ref{tab:mc_params}.


The emittances listed in Table \ref{tab:ext_params} are calculated as RMS emittances \cite{victor-cyklo} from the distributions in the phase space of particles.
\begin{equation}
\varepsilon_{rms,x}=4\sqrt{\overline{x^2} \cdot \overline{x'^2}- \overline{x \cdot x'}},
\label{eq:emitt}
\end{equation}

\noindent where $\overline{x^2}$ is the particle's position variance, $\overline{x'^2}$ is the variance of the particle's angle and $\overline{x \cdot x'}$ represents a correlation of the angle and position  of particles in the beam \cite{Buon}. The four dimensional transverse phase space ($x, x', z, z'$) is represented by a four dimensional emittance $\varepsilon_{rms,4D} = \varepsilon_{x} \varepsilon_{z}$.

\begin{table*}
\centering
\caption{Properties of the deflectors and magnetic channels optimized for the 37\,MeV proton mode and the $LE_2$ extraction mode. Values for the $PE$ extraction mode are shown in parentheses where appropriate.}
\begin{tabular}{@{}lllll@{}}\toprule
Parameter & Deflector D0 & Deflector D1 & MC 1 & MC 2 \\
\hline
    Start at  azimuth ($^\circ$) & 182$^\circ$ & 210$^\circ$ / 228$^\circ$ / 250$^\circ$ & 272$^\circ$ & 305$^\circ$  \\
    Azimuthal length ($^\circ$) & 22$^\circ$ & 18$^\circ$ \ / 22$^\circ$ \, / 18$^\circ$    & 23$^\circ$ & 55$^\circ$ \\
    Septum bending radius for $LE_2$ (mm) & 534  & 638 / 617 / 546  & 559  & 933   \\
    Septum bending radius for $PE$ (mm) & 545 & 639 / 619 / 543  & 582  & 992  \\
    Bending field in the center (kV/cm, T) & $-125$ ($-95$)& $-122$ ($-95$)  & 0 & 0  \\
    Bending field gradient (kV/cm$^2$, T/m) & $0$ &  $30$ (15) & $10$ (8) & $14$ (10) \\
    Radial gap length (mm) & 4 & 6 & 12 & 12 \\
\bottomrule
\end{tabular}
\label{tab:mc_params}
\end{table*}

\begin{figure}
    \centering
\begin{subfigure}[b]{0.47\textwidth}
     \centering
    {\includegraphics[width=\textwidth]{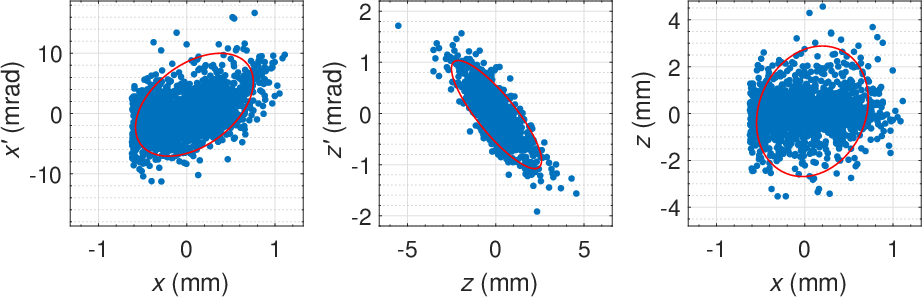}}
    \caption{Deflector D1 entrance at 183$^\circ$ and radius 524\,mm.}
    \label{subfig:ellipse_d0}
    \vspace{0.4cm}
\end{subfigure}
 \begin{subfigure}[b]{0.47\textwidth}
     \centering
    {\includegraphics[width=\textwidth]{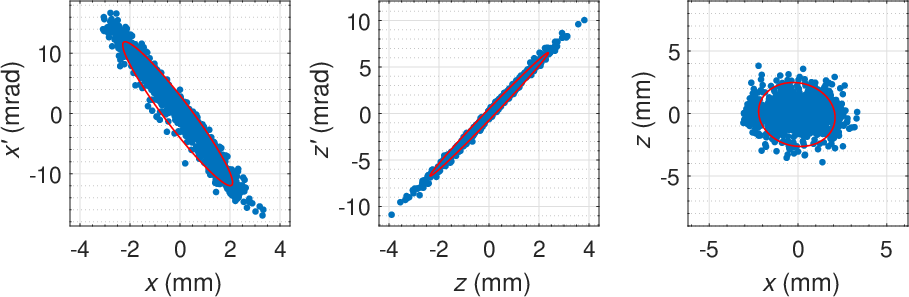}}
    \caption{Extraction system exit at 392$^\circ$ and radius 710\,mm.}
    \label{subfig:ellipse_ep}
\end{subfigure}
    \caption{Beam phase ellipses at the entrance and at the exit of the extraction system for the 37\,MeV proton mode and the $LE_2$ extraction system setting.}
    \label{fig:beam_ellipses}
\end{figure}

\begin{figure}
    \centering
\begin{subfigure}[b]{0.46\textwidth}
     \centering
    {\includegraphics[width=\textwidth]{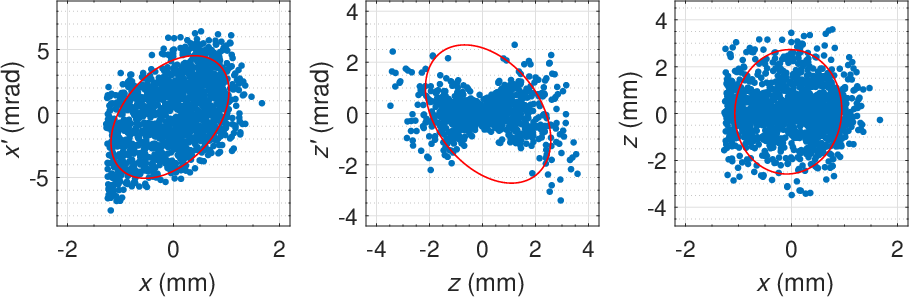}}
    \caption{Deflector D1 entrance at 183$^\circ$, radius 524\, mm.}
    \label{subfig:ellipse_d0_p}
    \vspace{0.4cm}
\end{subfigure}
 \begin{subfigure}[b]{0.46\textwidth}
     \centering
    {\includegraphics[width=\textwidth]{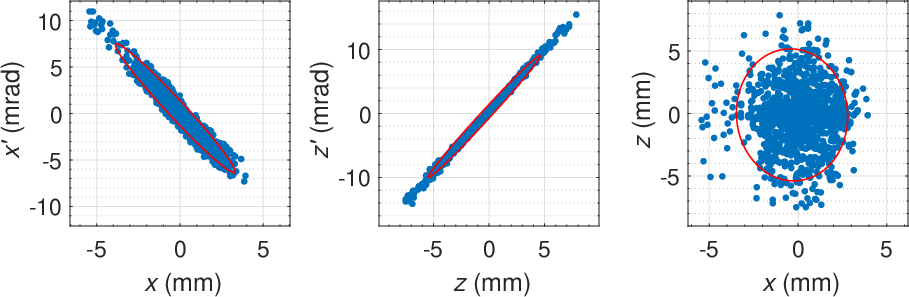}}
    \caption{Extraction system exit at 392$^\circ$ and radius 710\,mm.}
    \label{subfig:ellipse_ep_p}
\end{subfigure}
    \caption{Beam phase ellipses at the entrance and at the end of the extraction system for the 37\,MeV proton mode and the precessional extraction system setting.}
    \label{fig:beam_ellipses_prec}
\end{figure}

\begin{table}[ht]
\centering
\caption{Overall extraction efficiency $\eta_{\rm{T}}$ as obtained from Eq.\,\ref{eq:eff_tot}  and emittances of the 37\,MeV proton beam with RF phase size 15$^\circ$ extracted using different extraction modes. See text for further details.}
\begin{tabular}{@{}llll@{}}
Extraction mode & $PE$ & $LE_1$ & $LE_2$ \\
\hline
    $\varepsilon_{r,r'}$ ($\pi$\,mm\,mrad) & 5.4 & 3.1 & 5.0 \\
    $\varepsilon_{z,z'}$ ($\pi$\,mm\,mrad)& 2.3 & 1.0 & 1.0  \\
    $\varepsilon_{rms,4D}$ ($\pi$\,mm\,mrad)$^2$& 12.4 & 2.9 & 5.1  \\
    \hline
    Overall extr. efficiency  $\eta_{\rm{T}}$ & 35\% & 75\%\ & 70\%\\
\bottomrule
\end{tabular}
\label{tab:ext_params}
\end{table}



\section{Discussion}
\label{sec:discussion}

By optimizing the positions and shape of the individual extraction elements, it was possible to achieve the extraction efficiency $\eta$ higher than 70\% in all investigated extraction methods. From the performed simulations, it seems that the extraction efficiency is mainly determined by the losses on the septum of the deflector D0, which comes into direct contact with the accelerated beam.

The extraction efficiency for $LE_1$ reaches 80\%. This value results from 85\% transparency of the deflector D0, 95\% transparency of the deflector D1, and lossless magnetic channels. However, for this variant, the electric field in the deflector D1 reaches 150\,kV/cm, which can be difficult to achieve in practice. As shown in Fig.\,\ref{fig:d1_524_efield_map}, for a given electric field in the center of the deflector D1, the electric field at the edges of the profiled electrode is about 30\% higher. In general, the value of the electric field 150\,kV/cm in the centre of the deflector D1 can probably be considered too high. For this reason, the $LE_1$ extraction method is unsuitable for the highest energies of protons and helium-3 and can only be considered for intermediate and lower energies.

In the case of the second extraction variant $LE_2$, the electric fields in the deflector D1 are lower and reach about 120\,kV/cm. The lower field value is given mainly due to the drift of the orbital center towards the azimuth 270$^\circ$ when compared to the $LE_1$ variant, see Fig.\,\ref{subfig:center_le}. A larger shift of the orbit in this direction and further lowering the D1 voltage turns out not to be optimal, as the lower orbit separation is achieved at the entrance to the deflector D0 at azimuth 182$^\circ$, and the losses on the D0 septum increase.  Characteristics properties of the LE$_1$ and LE$_2$ extractions modes are compared in Table \ref{tab:hc_params}, where one can see that $LE_2$ has lower orbit separation (1.3\,mm vs. 1.5\,mm) and 
lower transmission in the D0 deflector (81\% vs. 85\%).


\subsection{Radial bunch size}

The radial size of the beam at the entrance of the deflector D0 is below 4\,mm for the proton modes and the helium-3 mode when using the linear resonance extraction. However, for the helium-3 mode with the precessional extraction, as well as for deuterium and alpha particles in the $LE$ modes, the radial beam size can be up to 6\,mm. Therefore, the new deflector D0 needs to have a variable gap size, ideally adjustable between 3 to 6\,mm. This approach is consistent with the current practice for the U-120M deflectors, where the gap size is optimized for specific experimental conditions.

The second deflector D1 has optimized its electric field gradient through a precisely shaped profile of the septum and the high-voltage electrode. Without this gradient, the radial width of the beam would be approximately 14\,mm at azimuth 270$^\circ$. The optimized deflector gradient for the 37\,MeV proton $LE_1$ mode is 26\,kV/cm$^2$, keeping the bunch size below 6\,mm to fit into the 6\,mm gap of the D1 deflector. However, for the $LE_2$ mode, the gradient is insufficient to fully focus the beam into the gap. To limit the losses on the high-voltage electrode and the septum at the D1 output, the electric field gradient must be increased to 30\,kV/cm$^2$. In the 55\,MeV helium-3 mode with the $LE$ extraction, the optimal deflector D1 gradient is approximately 20\,kV/cm$^2$. For the helium-3 mode and the $PE$, where the bunch radial expansion is minimal, the D1 gradient is not necessary at all.

\subsection{Precessional extraction}

By employing a low first harmonic component $\Delta B_{1\mathrm{OHC}}$ with an amplitude of 3--6\,Gauss, generated by the new harmonic coils, the amplitudes of radial coherent oscillation can be regulated to remain comparable with the incoherent oscillation amplitudes, as recommended in Ref. \cite{Hagedoorn-studies}. Following this guideline allows us to pass the coupling resonance $\nu_r = 2\nu_z$ with restricted vertical growth and maintaining the vertical bunch dimension within the dee vertical clearance of 20\,mm. The azimuth of the maximal orbit separation is determined  by varying the phase of the $\Delta B_{1\mathrm{OHC}}$  component \cite{victor-cyklo}.

The RF phase size of the bunch is constrained to approximately 15$^\circ$ as the precession does not occur at the same radius for all RF phases, see Fig. \ref{subfig:pr_precess}. The bunch is extracted immediately after the first precession, although there is a possibility to let the beam pass the precession resonance twice and achieve greater turn separation \cite{Hagedoorn2, gordon}. Based on our experience with precession excitation simulations, it might be possible to extend the beam RF phase size to about 30--40$^\circ$ while keeping the extraction efficiency $\eta$ still above 60\%. However, further extension of the interval poses a problem because particles with RF phases on the edges of the RF phase interval are poorly centered, leading to precession at different radii. These parts of the bunch enter the first deflector at varying angles and are prone to being lost on the septum or on the high-voltage electrode. From the point of view of minimizing the losses during the extraction by the $PE$ method, it is a common practice to restrict the beam RF phase size to 15--20$^\circ$ \cite{victor-cyklo}.

In Table \ref{tab:ext_params}, the overall efficiency $\eta_T$ is calculated for the $PE$ with the bunch RF phase size widened to 40$^\circ$ and is half that of the $LE_2$. In practice, due to the narrower beam RF phase width, this value is approximately 24\% for the RF phase size 15$^\circ$, which puts this method at a disadvantage compared to the $LE$.

Nevertheless due to the acceleration of the beam to the region $\nu_r < 1$, the extraction radius and the final energy is higher, thus the required voltage in the D1 deflector can be reduced by about 10--20\% compared to both the $LE$ modes. The necessary electric fields in the deflector D1 for the $PE$ mode, as listed in Table \ref{tab:hc_params}, are only 95\,kV/cm for 37\,MeV protons and 78\,kV/cm for 55\,MeV helium-3, in comparison  with 122\,kV/cm and 98\,kV/cm required in case of the $LE_2$ mode.

The $PE$ deflector's electric fields are remarkably low also in comparison with the current extraction system, where 180\,kV/cm is necessary to extract helium-3 at 55\,MeV. Given the rather problematic experience on the U-120M cyclotron with deflectors' operation in regions above 120\,kV/cm, which often leads to frequent discharges and high dark current, the reduction of the electric field required to deflect the maximum energy beams is likely to be a decisive factor in the selection of a suitable extraction method.

\subsection{Linear resonance extraction}

Using a significantly larger $\Delta B_{1\mathrm{OHC}}$ component of about 50\,Gauss, the beam can be drifted toward the D0 septum and extracted before the $\nu_r = 1$ resonance. This allows us to avoid the passage through the coupling Walkinshaw resonance and preserve the transverse dimensions of the bunch entering the cyclotron's fringe field. This enables the extraction of a wider RF phase interval for helium-3 or the entire interval for protons. In current state of the simulations for the helium-3 mode, the lowest RF phases mostly contribute to the losses, therefore the bunch RF phases below $-50^\circ$ are restricted by the phase slits in the central region, as shown in Fig.\,\ref{subfig:res_free_bar}. For the 37\,MeV proton mode, the bunch is not modified by the phase slits and is accelerated over the full RF phase interval. Due to RF phase mixing \cite{Hagedoorn2}, the output beam energy resolution is worsened, and due to the lower extraction radius, the output energy is reduced by about 10\%, as shown in Table \ref{tab:hc_params}. Nevertheless, in some applications, the higher output intensity can be of high importance, and lower output energy and higher energy dispersion may be acceptable. 

Although the overall extraction efficiency of $LE$ methods is several times higher than with the $PE$, the increased deflector voltage requirements will not make the $LE$ methods easy to implement and the benefits of higher output intensities will likely only be available for the lower output energy modes of the cyclotron.  


\subsection{Septum position}

The optimal position of the septum in the $PE$ mode is determined by the radius, where the precession takes place. In case of the proton and helium-3 mode, this position differs by 5\,mm. For the $LE$ modes, the radius is chosen carefully to prevent the lower RF phase part of the bunch from expanding vertically. For example, for the helium-3 mode with the $LE_2$ extraction setting, the low RF phase part of the bunch starts to grow vertically at the radius 526\,mm and the full bunch is completely lost at 530\,mm. In the $LE_1$ mode, the same bunch can be accelerated up to a radius of 530\,mm and extracted without  vertical expansion up to RF phase $-50^\circ$. 

\subsection{Shape of the extraction elements}
Trajectories of ions in the region occupied by deflectors are very similar for all three  extraction modes. The comparison in Table\,\ref{tab:mc_params} shows that the parameters of the trajectories in the deflector D1 are practically identical for the $LE$ and $PE$ extraction modes, although the energy of the ion extracted by the $PE$ mode is almost 10\% higher, see Table\,\ref{tab:hc_params}. If the resulting shape of the new deflector D1 is optimized for the shape of the particle trajectory, it can be assumed that it can be operated in both $PE$ and $LE$ extraction modes without significant modifications, adjusting only its actual position.

The shape of the ion path in the region of the D0 deflector varies slightly and in order to maintain maximum extraction efficiency it will be advantageous to adapt the exact shape of this deflector to the specific extraction mode, as in the case of magnetic channels.

\subsection{Emittance growth}

Table \ref{tab:ext_params} presents emittances calculated in horizontal $\varepsilon_{x,x'}$ and vertical $\varepsilon_{z,z'}$ phase space using Eq. \ref{eq:emitt} and compares the three investigated extraction methods. The initial full RF phase bunch is restricted by the internal phase slits to the RF phase size of 15$^\circ$, see Fig.\,\ref{subfig:preces_bar}, and accelerated under the same conditions to radius approximately 400\,mm, where it starts to be influenced by the outer HCs. The bunch is then extracted using the different setting of the three extraction modes $PE$, $LE_1$, and $LE_2$. The emittance of the beam extracted using the $PE$ mode is degraded by passing the coupling resonance $\nu_r = 2\nu_z$, which leads to an increase in the vertical emittance compared to both $LE$ modes. The distorted shape of the phase plot in the vertical phase space ($z,z'$) is evident from a comparison of Fig.\,\ref{subfig:ellipse_d0_p} for the $PE$ mode and Fig.\,\ref{subfig:ellipse_d0} for the $LE_2$. In both $LE$ modes, the bunch vertical phase space shape is unaffected by the coupling resonance and the bunch is entering the extraction system with reasonably lower emittance resulting in a smaller area in the $EP$ plane into which the beam can be focused.

\subsection{Magnetic channels}

The magnetic channels  for the new extraction system are considered to be fully passive channels i.e. consisting of only iron focusing bars magnetized by the main cyclotron magnetic field. The magnetic channels (MCs) are intended to be radially focusing with zero added magnetic field in their center i.e. without a bending magnetic field component \cite{Hagedoorn2}.

The basic design of the MCs was carried out first by designing a 2D profile using the analytical formulas given in Ref. \cite{victor-cyklo}. The obtained shape of the bars was subsequently used for the construction of realistic 3D CAD models of low carbon steel 1010 rods and the MCs magnetic field was then calculated in 3DS Opera \cite{CST}.  The beam dynamics simulations involving the simulated realistic MCs show great influence of their first harmonic magnetic field component to the cyclotron main magnetic field resulting in a significant RF phase slip of the accelerated bunch and lower extraction efficiency. Compensation of the disturbing first harmonic component is challenging and is planned to be done via a magnetic channels compensator \cite{gordon} placed azimuthally 180$^\circ$ from the focusing MCs. In Fig.\,\ref{fig:new_ext}, the magnetic channel compensator is denoted as MCC and is considered as a passive ferromagnetic structure, i.e. an iron bar fixed at a suitable place between the magnet poles. Compensation of the magnetic field average component can be done by means of altering the trim-coil's currents.


\section{Conclusion}

During the 2022 cyclotron shutdown caused by a short circuit in one of the trim-coils, new harmonic coils were installed in the acceleration chamber. Overcoming challenges related to limited vertical space, the coils were designed and constructed for placement at an extraction radius of 50\,cm.

At the beginning of the shutdown, cyclotron magnetic field measurements revealed a substantial component of the first harmonic disturbance. After the repair of the trim-coil, magnetic field measurements were made both for the new harmonic coils and for the entire cyclotron field. Precise adjustments to the position of the acceleration chamber between the poles of the magnet enabled to lower the first harmonic perturbation component to appro\-xi\-mate\-ly \\2\,Gauss at the extraction radius. The  measured maps of the field produced by the new harmonic coils were used to analyze the extraction of positive proton and helium-3 beams within the highest magnetic field of the cyclotron.

This study is focused on the modification of the existing extraction system. In this revised design, the role of the exciter in building coherent oscillations is taken over by the new harmonic coils and the exciter itself is replaced by a short electrostatic deflector D0 located in front of the old first deflector D1. The old deflector D1 was completely re-designed and the old second and the third electrostatic deflectors are substituted with passive magnetic channels.


The performed simulations showed that by carefully adjusting the currents in the new harmonic coils, it is possible to excite two different modes of resonant separation of orbits at the extraction radius. With a first harmonic component of approximately 3--4\,Gauss, the beam can be effectively extracted using the precession method. On the other hand, exciting a first harmonic component of around 50\,Gauss enables the use of the integer or linear resonance method. 

For the maximal extrac\-ted beam intensity, the linear resonance enables extraction across a wide range of RF phases. Simulated extraction efficiency in this scenario is within the range of 70--75\%. Achieving the high overall extraction efficiency comes at the cost of reduced output energy and the necessity for higher voltages on the deflectors due to extraction from a smaller radius.

The precessional extraction method proves more suitable for applications demanding higher output energy, achieving approximately 10\% higher output energy compared to the previous method. This increase results from the acceleration of the beam up to the fringe field of the cyclotron, where the radial frequency $\nu_r$ drops below unity. This method offers the advantage of reducing the electric field of the deflectors due to the higher extraction radius and the increased energy of the extracted beam. 

In the case of precession method the electric fields in the deflectors are  by approximately 10--20\% lower then with the linear resonance extraction and by around 50\% lower compared to the current U-120M extraction system. The precessional method has also limitations, particularly in terms of the range of RF phases. The precession of the beam does not occur for all RF phases at the same radius, leading to constraints on the output intensity. To minimize losses on the first deflector, simulations in this study emulate the use of phase slits and limit the range of RF phases to 15$^\circ$, resulting in an extraction efficiency above 70\%. Nevertheless, the overall extraction efficiency, which corresponds to the output beam intensity, is significantly lower than that of the linear resonance method.

Given the above, the optimal strategy would be to have possibility to use both extraction methods and to take advantage of the strengths of each.  However, this would mean constructing two slightly different versions of the extraction system and using them as needed, tailored to the requirements of a particular experiment.

In case that it would be not possible to construct two versions of the extraction system, a comparative analysis of the two resonance methods suggests that the precession method is the more advantageous for implementation,  because it can achieve higher energies and it operates the electrostatic deflectors at lower voltages. Moreover, the energy range of the U-120M cyclotron would be extended. Despite the lower intensity of the total output beam compared to the linear resonance method, the next steps will be to implement the simulation results presented in this work and to construct a new extraction system based on the precession principle.

The advantage of the linear resonance method, consisting in smaller beam emittance and several times higher intensity of the output beam, may find important applications in experiments with a small effective cross section, e.g. in the development of the new radiopharmaceuticals. For the lower energies of the accelerated beam, where the disadvantage of higher deflector voltages does not apply, the linear resonance method offers the possibility of extracting a significant part of the internal beam.

\section*{Acknowledgement}

The authors would like to thank Dr. Filip Křížek for his careful reading of the text and his insightful comments on the edits.


\bibliographystyle{unsrtnat} %
\bibliography{bibi}

\end{document}